\documentclass[runningheads]{llncs}
\usepackage[T1]{fontenc}
\usepackage{graphicx}
\usepackage{cite}
\usepackage{hyperref}
\usepackage{xcolor}
\usepackage{array}
\usepackage{cite}
\usepackage{multirow}
\newcolumntype{?}{!{\vrule width 2pt}}

\begin{document}

\title{Improving Generalization Capability of Deep Learning-Based Nuclei Instance Segmentation by Non-deterministic Train Time and Deterministic Test Time Stain Normalization \thanks{Partially supported by the Austrian Research Promotion Agency (FFG), No. 872636.}}

\titlerunning{Non-deterministic Normalization for Nuclei Instance Segmentation}
\author{Amirreza Mahbod\inst{1} \and
	Georg Dorffner \inst{2} \and
	Isabella Ellinger \inst{3} \and 
	Ramona Woitek \inst{1} \and 
	Sepideh Hatamikia \inst{1,4}
}

\authorrunning{A. Mahbod et al.}

\institute{Research Center for Medical Image Analysis and Artificial Intelligence, Department of Medicine, Danube Private University, Krems an der Donau, Austria \and
Institute of Artificial Intelligence, Medical University of Vienna, Vienna, Austria \and 
Institute for Pathophysiology and Allergy Research, Medical University of Vienna, Vienna, Austria \and
Austrian Center for Medical Innovation and Technology, Wiener Neustadt, Austria
}


\maketitle              

\begin{abstract}
With the advent of digital pathology and microscopic systems that can scan and save whole slide histological images automatically, there is a growing trend to use computerized methods to analyze acquired images. Among different histopathological image analysis tasks, nuclei instance segmentation plays a fundamental role in a wide range of clinical and research applications. While many semi- and fully-automatic computerized methods have been proposed for nuclei instance segmentation, deep learning (DL)-based approaches have been shown to deliver the best performances. However, the performance of such approaches usually degrades when tested on unseen datasets. 

In this work, we propose a novel method to improve the generalization capability of a DL-based automatic segmentation approach. Besides utilizing one of the state-of-the-art DL-based models as a baseline, our method incorporates non-deterministic train time and deterministic test time stain normalization, and ensembling to boost the segmentation performance. We trained the model with one single training set and evaluated its segmentation performance on seven test datasets. Our results show that the proposed method provides up to 4.9\%, 5.4\%, and 5.9\% better average performance in segmenting nuclei based on Dice score, aggregated Jaccard index, and panoptic quality score, respectively, compared to the baseline segmentation model.


\end{abstract}

\newpage

\section{Introduction}
The evaluation of histopathological images by experts remains an integral part of the diagnostic routine of many human diseases~\cite{irshad2014methods}. An essential element of this process is the inspection of the appearance, morphology, and density of cells, which is subsequently used, for example, to diagnose different types of cancer or to assess the progression of certain diseases~\cite{https://doi.org/10.1111/tbj.13728,fischer2020nuclear, Basu2023}. Another important aspect in this context is the shape and structure of the cell nuclei~\cite{skinner2017nuclear}. Nuclear segmentation -- the task of finding all individual nuclei in digitized histological images -- is therefore a key feature of many automated pathology frameworks, as it enables the subsequent extraction of important information, including cell count or quantities related to the shape and structure of the nuclei~\cite{diagnostics13142379}. Since automated pathology frameworks can improve diagnostic accuracy, reduce evaluation time, and create more efficient workflows, there has been an increased effort to automate nuclei segmentation with the goal of achieving more robust and objective segmentation results~\cite{7373572,monuseg,mahbod2021cryonuseg_org}. Many computerized methods have been proposed for unsupervised, semi-supervised, and fully-supervised nuclei segmentation ranging from standard image processing techniques to advanced machine learning (ML) and deep learning (DL) algorithms~\cite{HOLLANDI2022295,meijering2012cell}. However, among these methods, supervised DL algorithms, such as convolutional neural networks (CNN), have achieved the best performances~\cite{HOLLANDI2022295,anwar2018medical}. Although there are many different nuclei instance segmentation algorithms based on CNNs, the state-of-the-art strategies can be broadly categorized into three different categories. There are detection-based methods such as adapted or improved Mask R-CNN~\cite{pmlr-v156-bancher21a,johnson2018adapting}, ternary segmentation models such as deep contour-aware networks (DCAN) or Kumar et al. method~\cite{Kumar2017,chen2017dcan} and distance-based methods such as Hover-Net~\cite{graham2019hover}, two-stage U-Net algorithms~\cite{10.1007/978-3-030-23937-4_9} or dual decoder U-Net-based (DDU-Net) model~\cite{10.3389/fmed.2022.978146}. Additionally, these approaches can be jointly used in order to boost the segmentation performance~\cite{pmlr-v156-bancher21a,8759574}. Furthermore, considering the recent advances of large language models, vision transformer-based architectures have also been utilized in the encoder-decoder-based models for various histological image analysis tasks, including nuclei instance segmentation~\cite{10190115, SHAMSHAD2023102802}. Comprehensive overviews of state-of-the-art methodologies for nuclei segmentation can be found in the respective studies~\cite{Basu2023,HOLLANDI2022295,meijering2012cell, 10190115}.


Despite the fact that supervised CNNs achieve excellent nuclei segmentation performances for individual datasets, the CNN performance usually diminishes on external or unseen test data, because histological images are usually acquired under different settings~\cite{tellez2019quantifying, SALVI2021104129}. Specifically, there are many sources of variations in the appearance of cells acquired in different labs, including color variations caused by minor stain deviations, different organs, or different image acquisition devices. All these variations in image acquisition settings pose a challenge known as domain shift, the differences between the source domain (the training data) and the target domain (the unseen test data). This challenge arises because the model, during training, learns to recognize patterns and features that are specific to the training domain but when faced with the test data from a different domain, the model may encounter unfamiliar variations and struggle to generalize effectively~\cite{VASILJEVIC2023110780, jahanifar2023domain}. Thus, creating an algorithm that performs well on all datasets is a challenging task. Although many efforts were made to overcome these obstacles and improve the generalization power of DL-based methods,  there is still a lack of robust algorithms with acceptable nuclei segmentation performance for unseen test datasets~\cite{graham2019hover,mahbod2022deep}. 

Normalization-based and augmentation-based approaches are the most well-known applied methods to improve generalization~\cite{tellez2019quantifying,10.3389/fbioe.2019.00300}.

Normalization algorithms focus on reducing the variability of input images from different sources. They often match certain properties of the target/input image to those of a reference image~\cite{tellez2019quantifying,10.3389/fbioe.2019.00300}. Various methods have been proposed to perform normalization in histological images. These include classical image processing techniques such as histogram matching~\cite{reinhard2001color}, stain-based normalization such as  Macenko et al. method~\cite{macenko2009method} or Vahadane et al. method~\cite{7164042}, and neural network-based approaches such as cycle generative adversarial networks (CGAN) or HistoStarGAN~\cite{8237506, VASILJEVIC2023110780}. Classical techniques such as Reinhard et al.~\cite{reinhard2001color} are not originally designed for computational pathology and can introduce undesirable artifacts in histological images~\cite{khan2014nonlinear}. Stain-based normalization methods are initially designed for matching source and target domain in histological images, and they have been shown to keep the structural details while altering the stain matrix of source images~\cite{jahanifar2023domain}. While the application of these methods has led to improved image analysis performance in some studies~\cite{SALVI2021104129, Jung2019}, a number of other studies have shown adverse effects or no effects when using these techniques on classification or segmentation performances in Hematoxylin \& Eosin (H\&E)-stained histological images especially when combined with ML-based or DL-based approaches~\cite{tellez2019quantifying, aresta2019bach, 10.1007/978-3-031-21014-3_26, Voon2023}.  GAN-based approaches have shown excellent performance in domain translation where there is a large gap between the source and target domain e.g., changing the staining type from H\&E to immunohistochemistry~\cite{8237506}. Although such approaches generally generate visually plausible image-to-image translation, they are still prone to create hallucinative image features, and they are quite sensitive to the model architecture and training procedure. Thus, their application for some tasks, including specific image segmentation in histological images, has shown to be limited~\cite{jahanifar2023domain, VASILJEVIC2022102420}.

Augmentation, on the other hand, exploits various geometrical or mathematical color transformations in order to introduce more variety during the training. Common techniques are rotations, mirroring, scaling, elastic deformations, or adding small perturbations to the channels in different color spaces~\cite{tellez2019quantifying, jahanifar2023domain}. Many widely recognized augmentation techniques have their roots in the domains of natural image classification or segmentation. However, they have demonstrated promising applications in medical image analysis, including histological image segmentation~\cite{jahanifar2023domain}.
  
Numerous valuable efforts have been dedicated to enhancing generalization in computational pathology. However, some studies have faced constraints due to the use of limited test sets or a restricted number of tissues/organs for assessing generalization capabilities~\cite{haq2023nusegda, 10.1007/978-3-031-16449-1_68, 9622821, 9156759, VASILJEVIC2023110780}. Additionally, certain investigations have simultaneously addressed multiple tasks, such as classification and segmentation~\cite{li2023laplacian}, while others have focused on proposing solutions for distinct tasks, such as virtual stain transfer in histological images~\cite{VASILJEVIC2022102420, VASILJEVIC2023110780, WANG2023107768}.

In this work, we explicitly focus on improving the generalization power for the nuclei instance segmentation task. We proposed and developed a hybrid approach by combining the normalization and augmentation techniques in the nuclei instance segmentation workflow. We use DDU-Net~\cite{10.3389/fmed.2022.978146}, one of the state-of-the-art DL-based nuclei instance segmentation models, as the baseline. For training, we incorporate non-deterministic stain normalization based on the Macenko et al. method. Instead of using a single reference image and performing normalization as an offline pre-processing, we select multiple reference image candidates from different organs. We normalize input images randomly as an online augmentation during the training phase. In the inference phase, besides using morphological test time augmentation, we also apply a deterministic test time stain normalization strategy followed by an ensembling step to create the final segmentation output. In our experiments, we used one single training set (including images from seven distinct organs/tissues) and evaluated the performance of seven test sets (including images from 40 organs/tissues). Our results on all test datasets show that the proposed method has improved nuclei instance segmentation performance compared to the baseline model. 

The main contributions can be summarized as follows:
\begin{itemize}
	\renewcommand{\labelitemi}{$\bullet$}
	\item  Integrating a novel non-deterministic stain normalization in the training procedure of DDU-Net. 
	\item   Applying novel deterministic weighted test time normalization in the inference phase. 
	\item  Utilizing best practices in the test phase including morphological test time augmentation and model ensembling.    
	\item  Conducting extensive experiments to demonstrate consistent improved nuclei instance segmentation performance on seven different test datasets compared to the baseline segmentation model.
\end{itemize}


\section{Materials and Methods}
\subsection{Datasets}
In our study, we used one single dataset for training and multiple datasets for testing to evaluate the generalization capability of our proposed approach (i.e., one single trained model was tested on multiple unseen images from various test datasets). The training was done on the training set from the MoNuSeg challenge~\cite{monuseg}, as it encompasses a large variety of nuclei from different organs and has been widely used as a benchmark dataset in previous studies~\cite{graham2019hover, ZHAO2020101786, 10.3389/fmed.2022.978146}. It includes a total of 21,623 nuclei, found in 30 ($1000\times1000$ pixels) H\&E-stained image patches extracted from whole slide images from the cancer genome atlas (TCGA) repository\footnote{\url{https://portal.gdc.cancer.gov/repository}}. This dataset includes images from seven organs, namely liver (6 images), breast (6 images), kidney (6 images), bladder (2 images), prostate (6 images), stomach (2 images), and colon (2 images). For testing,  we utilized seven test datasets. We used MoNuSeg test data~\cite{monuseg}, TNBC~\cite{naylor2018segmentation}, CryoNuSeg~\cite{mahbod2021cryonuseg_org}, CPM-15~\cite{vu2019methods}, CPM-17~\cite{vu2019methods}, CoNSeP~\cite{graham2019hover} and NuInsSeg dataset~\cite{mahbod2023nuinsseg}, all include image patches with H\&E staining. Further depiction of the attributes of each dataset can be found in Table~\ref{datasets}.

\begin{table}
\centering
\caption{Summary of the used datasets for training (first row) and testing (other rows). Training and testing data contain images from 7 and 40 distinct organs/tissues, respectively. }\label{datasets}
\begin{tabular}{l|c|c|c|c}
\hline
Datasets &  \# Nuclei & \# Images & Image Size &  \# Organs/tissues\\
\hline
MoNuSeg Training Set &  21,623  & 30  & $1000\times 1000$                  & 7\\ \hline
MoNuSeg Test Set     &  7,223   & 14  & $1000\times 1000$                  & 9 \\
TNBC                 &  4,022   & 50  & $512\times 512$                    & 1 (breast)\\
CryoNuSeg            &  7,596   & 30  & $512\times 512$                    & 10\\
CPM-15               &  2,905   & 15  & various sizes                     & 2\\
CPM-17               &  7,570   & 32  & $500\times 500$,  $600\times 600$ & 4\\
CoNSeP               &  24,319  & 41  & $1000\times 1000$                  & 1 (colon)\\
NuInsSeg             &  30,698   & 665  & $512\times 512$                    & 31 \\
\hline
\end{tabular}
\end{table}

\subsection{Segmentation model}
\begin{figure}
\centering
\includegraphics[width=\textwidth]{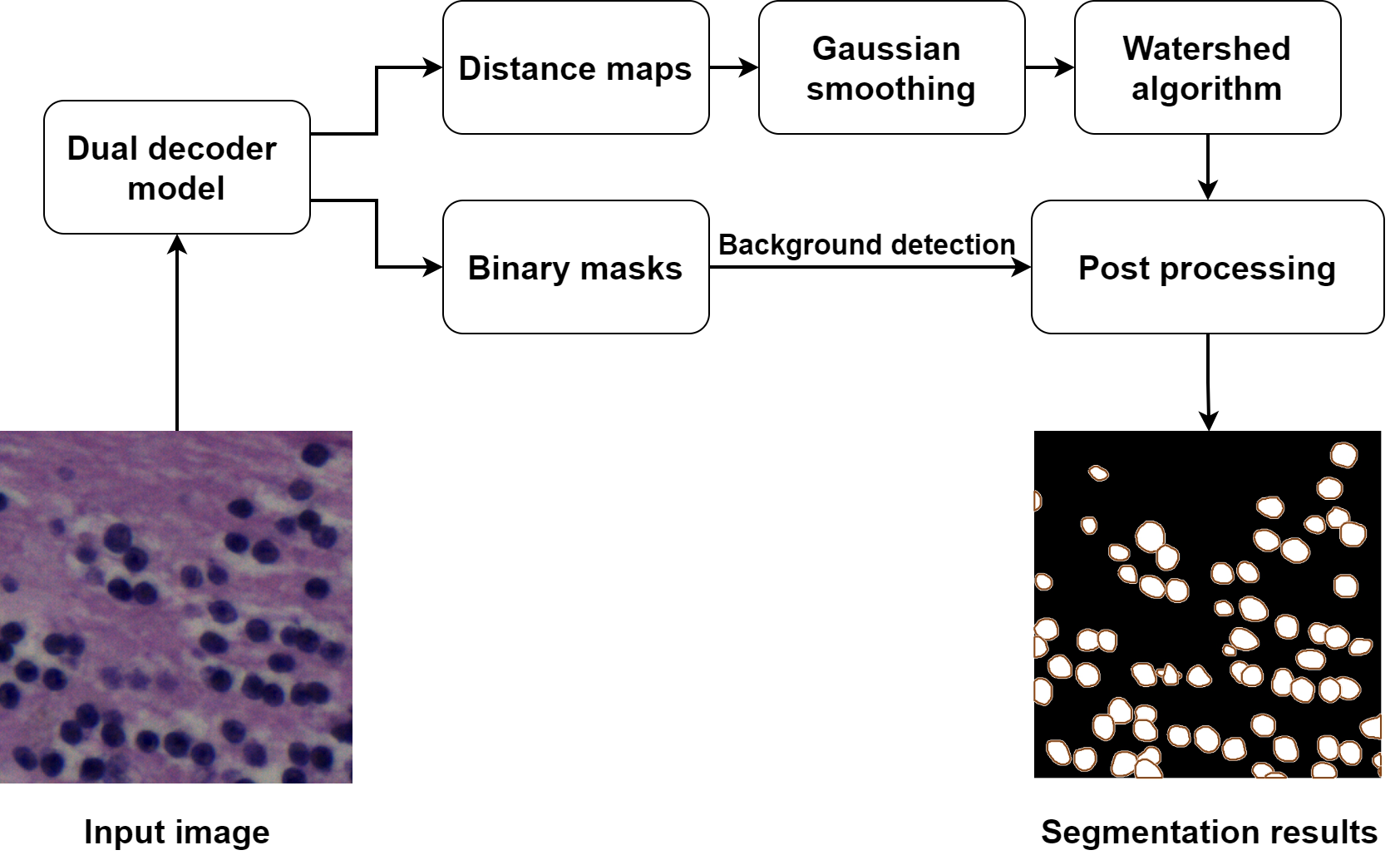}
\caption{The generic workflow of the DDU-Net~\cite{10.3389/fmed.2022.978146} for nuclei instance segmentation.} \label{seg_model}
\end{figure}

This work is based on the DDU-Net~\cite{10.3389/fmed.2022.978146} as the baseline nuclei instance segmentation model. This model has shown excellent performance in the nuclear segmentation task in various datasets (e.g., it achieved the first rank on the MoNuSAC post-challenge leaderboard for multi-organ nuclear segmentation and classification challenge~\cite{9745890, 9745980}). It uses a U-Net-alike encoder-decoder structure~\cite{Ronneberger2015}, such that the images are fed into a shared encoder path, whose intermediate results are consequently passed to two different decoder branches. These decoders are designed to predict nuclear pixels (first decoder) and distance maps (second decoder) of all instances in a given image. The shared encoder consists of convolutional, drop-out, and max pooling layers, while the decoders consist of convolutional, drop-out, and transpose convolutional layers. The only difference between the two decoders is the last layer, where a sigmoid activation function and a linear activation function are used for the first and second decoders, respectively. For the first decoder, a combined Dice loss and a binary cross-entropy loss were used, and for the second decoder, a mean square error loss function was utilized. The general workflow of the utilized DDU-Net is shown in Fig.~\ref{seg_model}. The results from decoders are post-processed using a Gaussian smoothing filter, a watershed algorithm, and morphological operations to form the final instance segmentation results. Further details about the model architecture and workflow can be found in the respective study~\cite{10.3389/fmed.2022.978146}.

\subsection{Non-deterministic stain normalization in training}
\label{Non-deterministic-norm}
In this work, we made use of the normalization algorithm introduced by Macenko et al.~\cite{macenko2009method}, which has been widely applied in the literature to normalize H\&E-stained histological images~\cite{GEORGE2020103954, Hameed2022}. This method is based on deconvoluting images to retrieve the pure hematoxylin (i.e., nuclei) and eosin (i.e., other cell parts) components. After applying the negative logarithm to convert the images into the optical density (OD) space, it is possible to model the stains with the following equation: 
\begin{equation}
	OD = S * V
\end{equation}

Where $S$ and $V$ are matrices containing the stain saturations (S) and the stain vectors (V). The stain vector, however, is not known and is hence approximated. Macenko et al. method~\cite{macenko2009method} achieves this by utilizing singular value decomposition. Afterward, the saturations of the single stains can be calculated using the formula above. While this is shown to preserve histological information~\cite{10.3389/fbioe.2019.00300}, the results rely on the sophisticated selection of the reference image, a step that is usually performed manually.

We introduced a new strategy for randomly applying the Macenko et al. method during training to improve the generalization capability, paired with an automated reference image selection algorithm, reducing the impact of reference image selection and creating more diversity in the training data. 

To select the reference images automatically, we used the MoNuSeg training data histograms. We calculated the mean intensity value for nuclei regions and background regions using the provided binary segmentation masks for the MoNuSeg training data. We sorted all images by the absolute difference between the mean tissue intensity and the mean nuclear intensity and selected the images with the largest differences between the means (one per organ). Table~\ref{difference} shows the differences for all images in the training set (selected reference images for each organ are shown in bold).

\begin{table}
	\centering
	\caption{The differences between the mean tissue intensity and the mean nuclear intensity for all training images (30 images) of the MoNuSeg dataset. The selected reference images are shown in bold.}\label{difference}
	\begin{tabular}{l|c|c?l|c|c}
		\hline
		Image ID &  Organ & diff.& Image ID &  Organ & diff.\\
		\hline
		TCGA-18-5592-01Z-00 & liver    & 29.82             & TCGA-AR-A1AS-01Z-00          & breast              & 66.36\\
		TCGA-49-4488-01Z-00 & liver    & 41.48             & TCGA-A7-A13E-01Z-00          & breast              & 81.31\\
		TCGA-50-5931-01Z-00 & liver    & 42.24             & \textbf{TCGA-A7-A13F-01Z-00} & \textbf{breast}     & \textbf{89.17}\\
		TCGA-38-6178-01Z-00 & liver    & 55.49             & TCGA-HE-7130-01Z-00          & kidney              & 48.46\\
		TCGA-21-5786-01Z-00 & liver    & 62.47             & TCGA-HE-7129-01Z-00          & kidney              &	59.03\\ 
		\textbf{TCGA-21-5784-01Z-00}   & \textbf{liver}    & \textbf{87.85}               & TCGA-HE-7128-01Z-00 & kidney & 71.01 \\
		TCGA-G9-6336-01Z-00 & prostate & 30.22             & TCGA-B0-5711-01Z-00          & kidney              & 92.40\\
		TCGA-G9-6363-01Z-00 & prostate & 30.63             & TCGA-B0-5698-01Z-00          & kidney              & 93.83\\
		TCGA-G9-6362-01Z-00 & prostate & 37.99             & \textbf{TCGA-B0-5710-01Z-00} & \textbf{kidney}     & \textbf{100.83}\\
		TCGA-G9-6348-01Z-00 & prostate & 44.62             & TCGA-RD-A8N9-01A-01          & stomach             & 108.58\\
		TCGA-CH-5767-01Z-00 & prostate & 47.62             & \textbf{TCGA-KB-A93J-01A-01} & \textbf{stomach}    & \textbf{109.36}\\
		\textbf{TCGA-G9-6356-01Z-00}   & \textbf{prostate} & \textbf{60.57}               & TCGA-AY-A8YK-01A-01 & colon  &	50.58\\
		TCGA-AR-A1AK-01Z-00 & breast   & 44.37             & \textbf{TCGA-NH-A8F7-01A-01} & \textbf{colon}      & \textbf{58.60}\\
		TCGA-E2-A1B5-01Z-00 & breast   & 57.81             & TCGA-G2-A2EK-01A-02          & bladder             &	74.07\\
		TCGA-E2-A14V-01Z-00 & breast   & 62.15             & \textbf{ TCGA-DK-A2I6-01A-01}& \textbf{bladder}    &	\textbf{84.13}\\
		\hline
	\end{tabular}
\end{table}

\begin{figure}[t!]
	\centering
	\begin{tabular}{cccc}
		Original image & Nuclei image & Tissue image & Histogram\\
		\includegraphics[width=3cm]{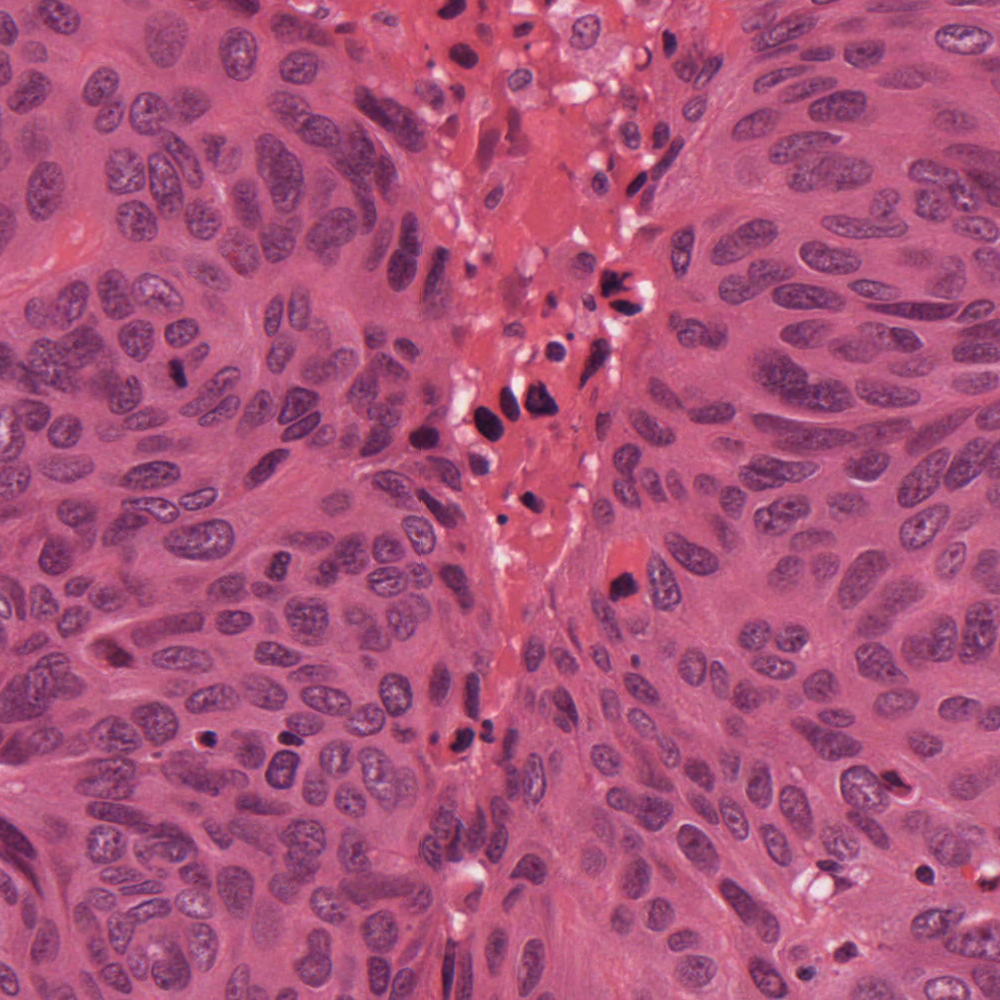} &
		\includegraphics[width=3cm]{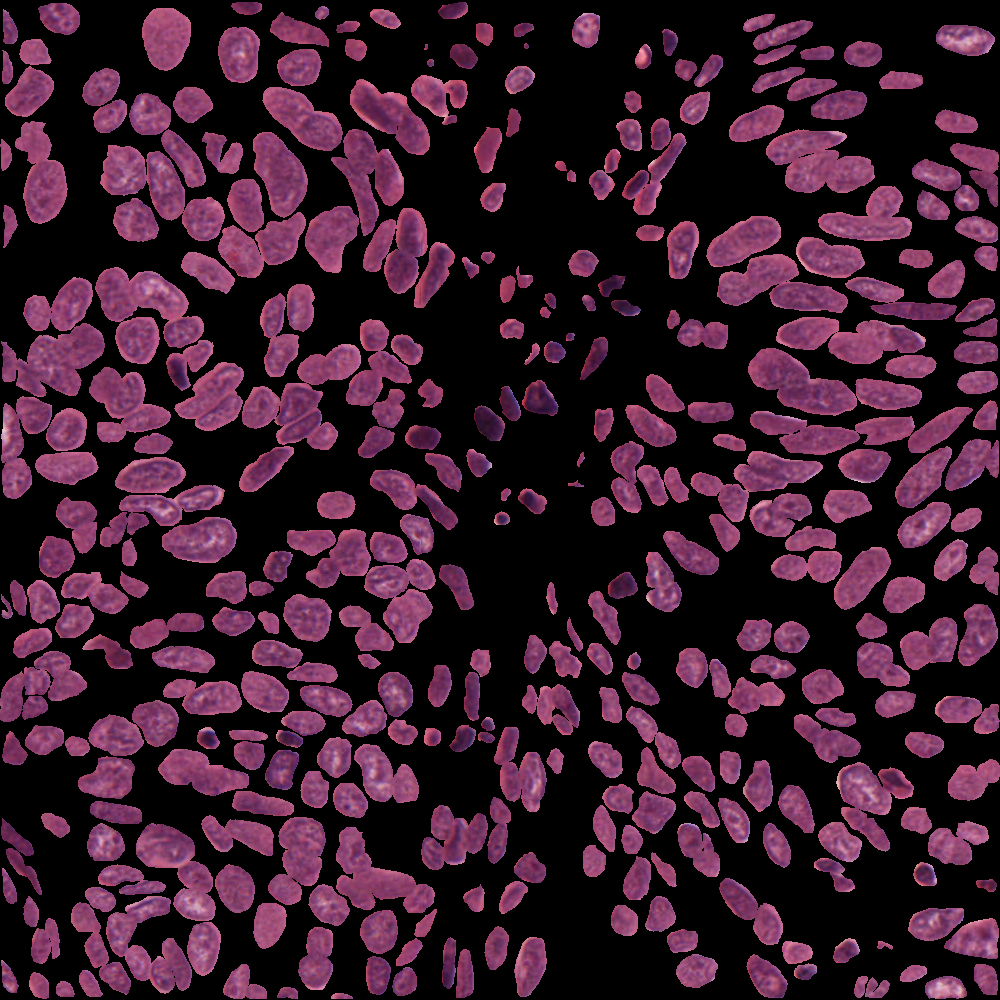} &
		\includegraphics[width=3cm]{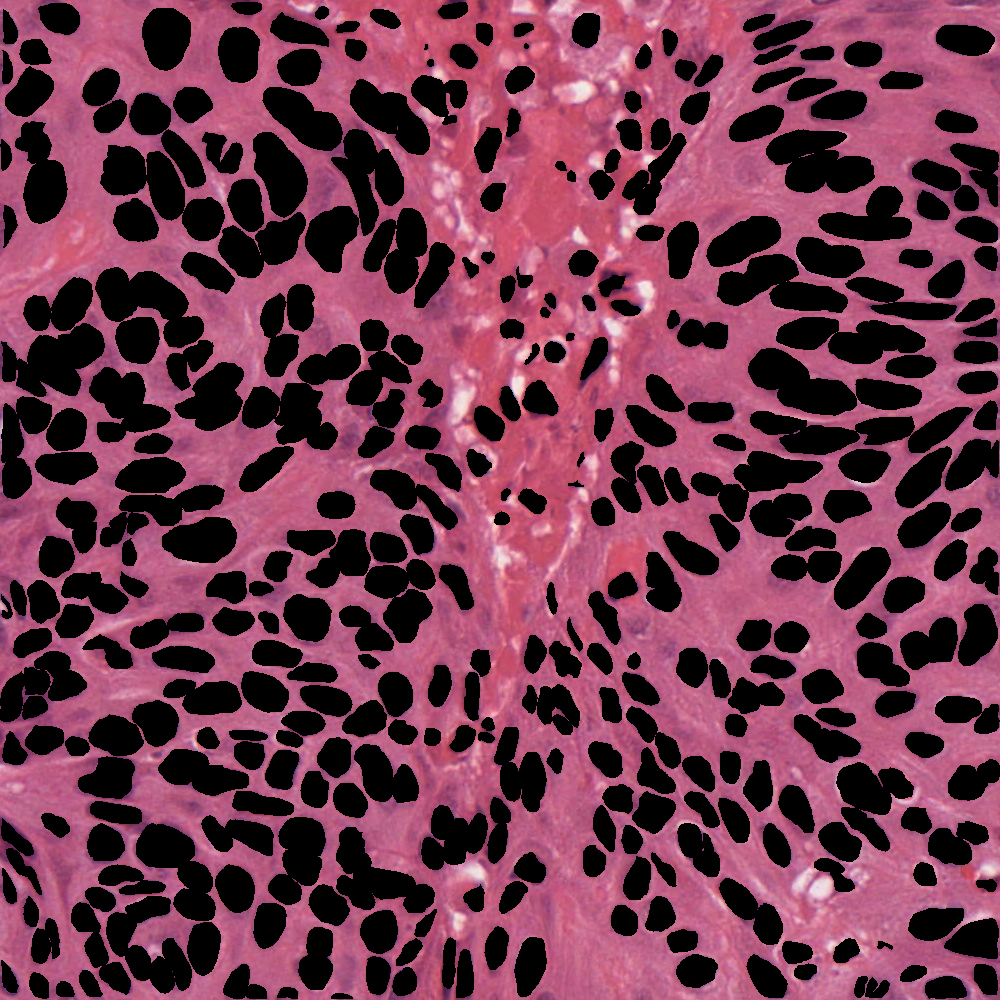} &
		\includegraphics[width=3cm]{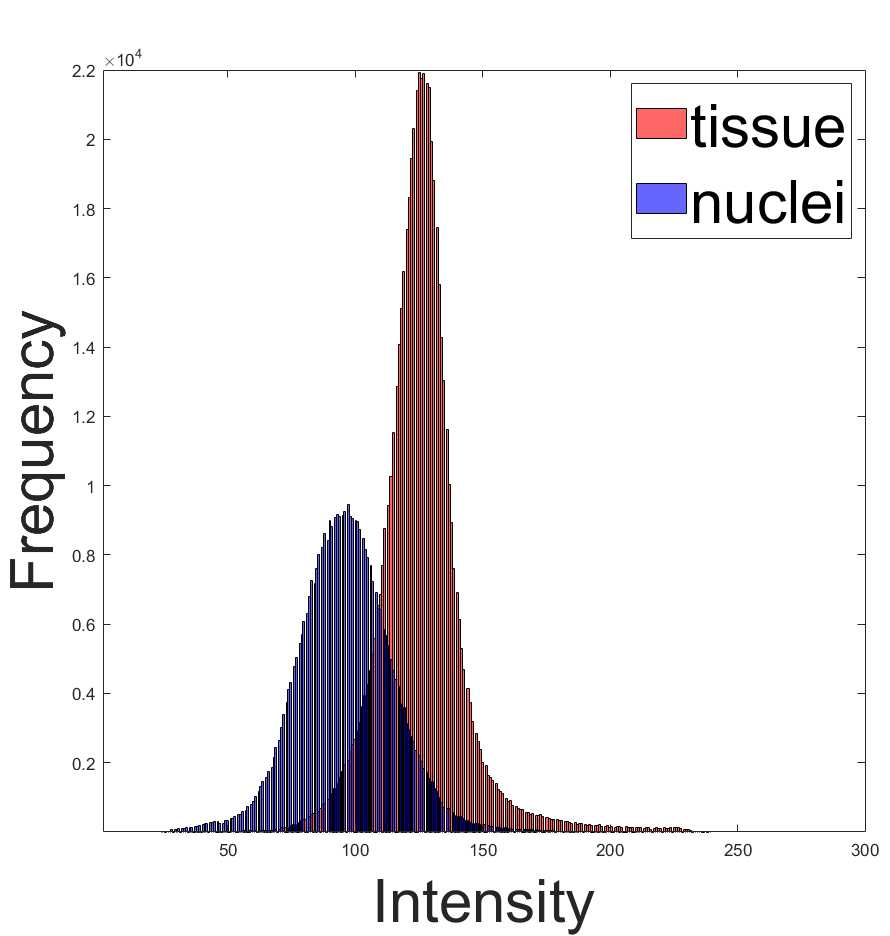} 
		\\
		\includegraphics[width=3cm]{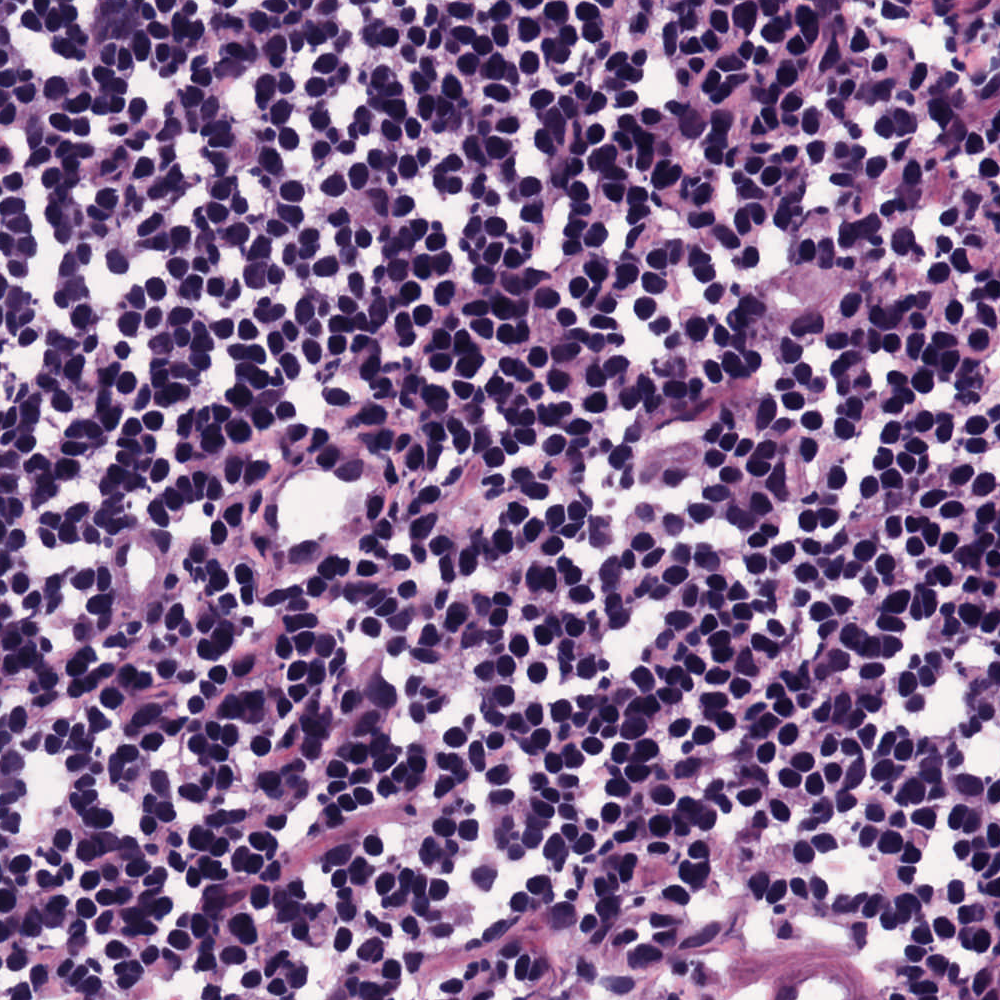} &
		\includegraphics[width=3cm]{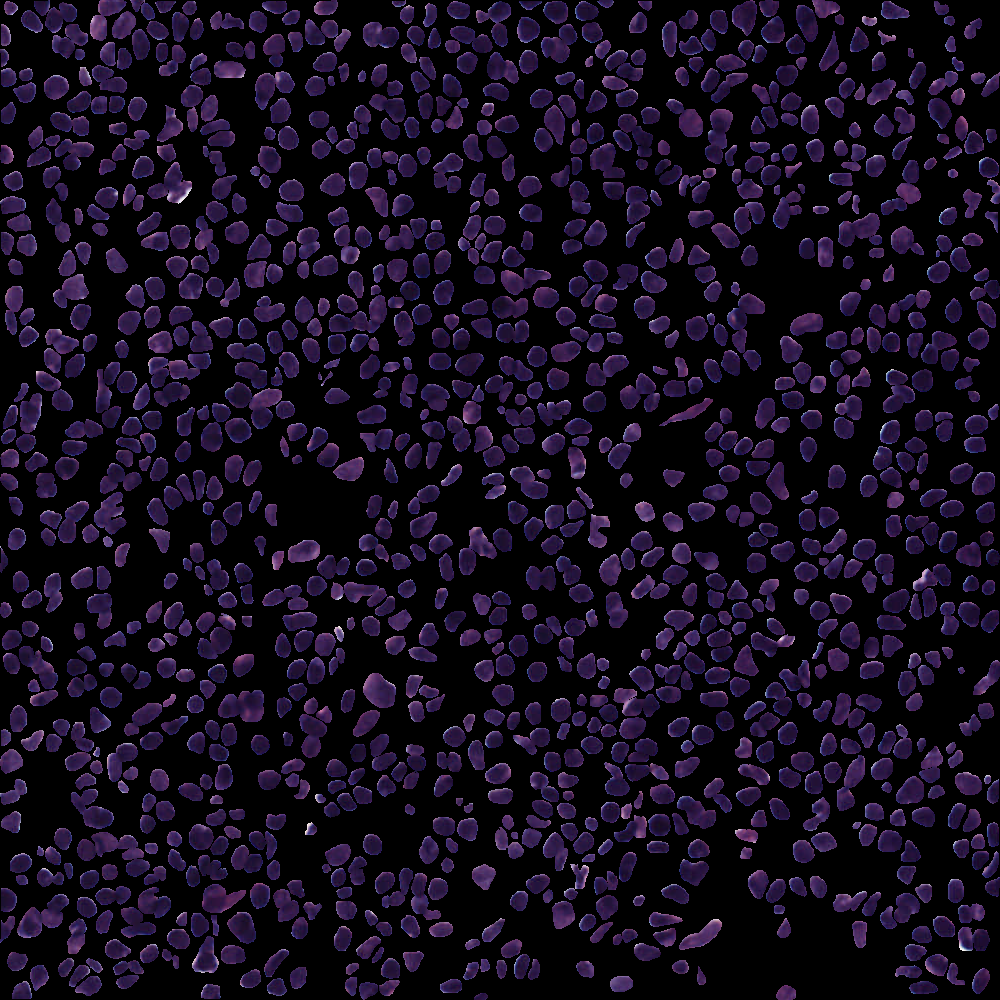} &
		\includegraphics[width=3cm]{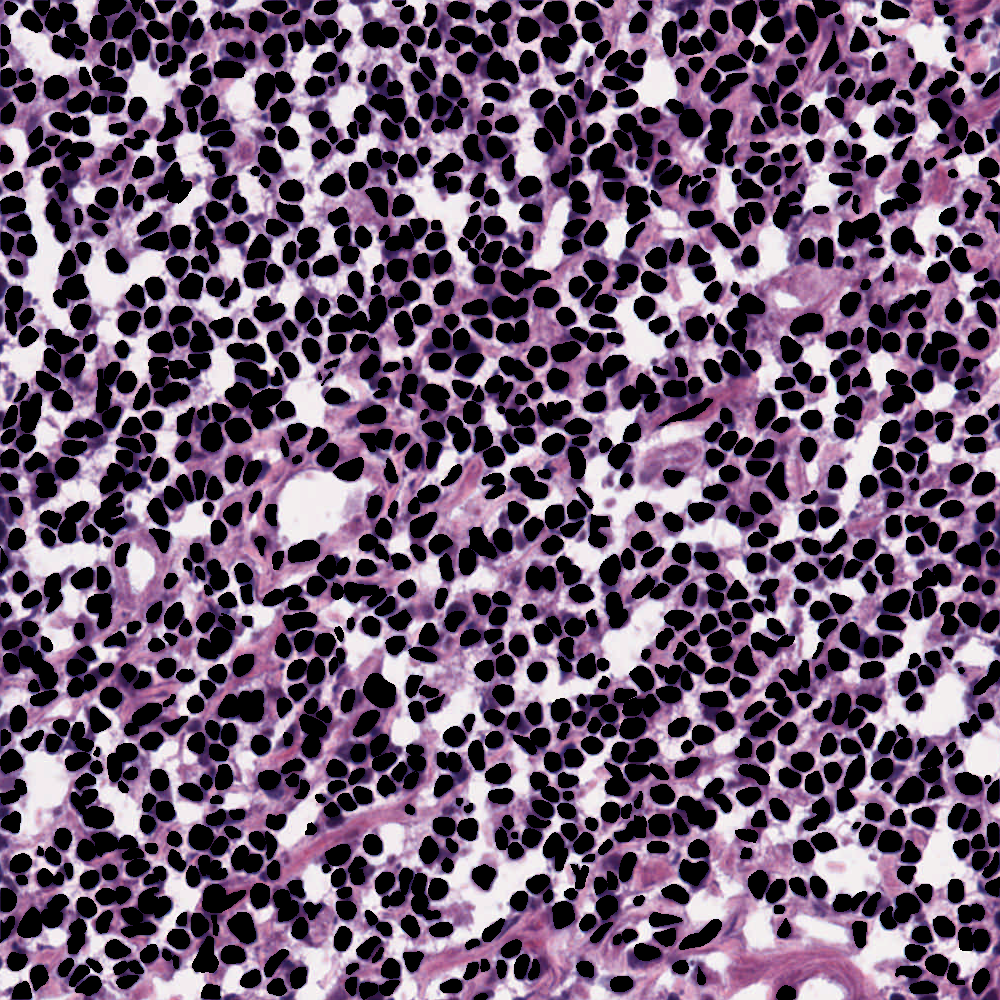} &
		\includegraphics[width=3cm]{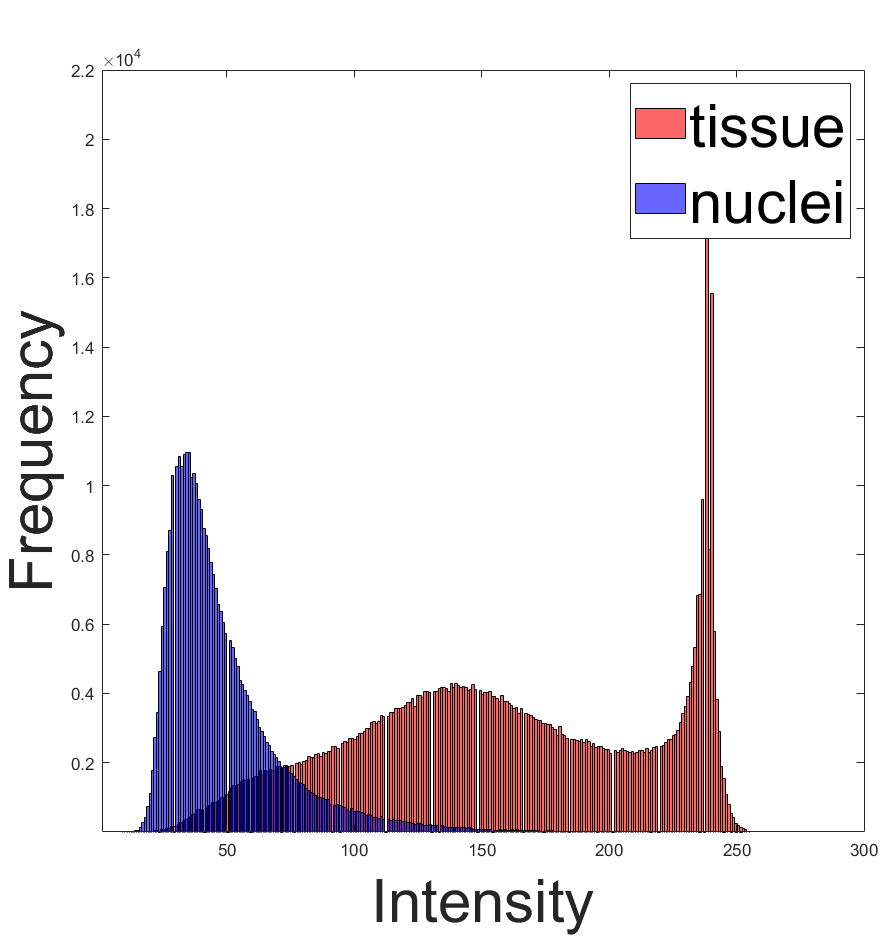} 
		
	\end{tabular}
	\caption{Examples of a non-selected (first row) and a selected (second row) reference images. }
	\label{fig:histogram}
\end{figure}

\begin{figure}[t!]
	\centering
	\begin{tabular}{ccccccc}
		Liver & Breast  & Kidney & Bladder & Prostate & Stomach & Colon\\
		\includegraphics[width=1.6cm]{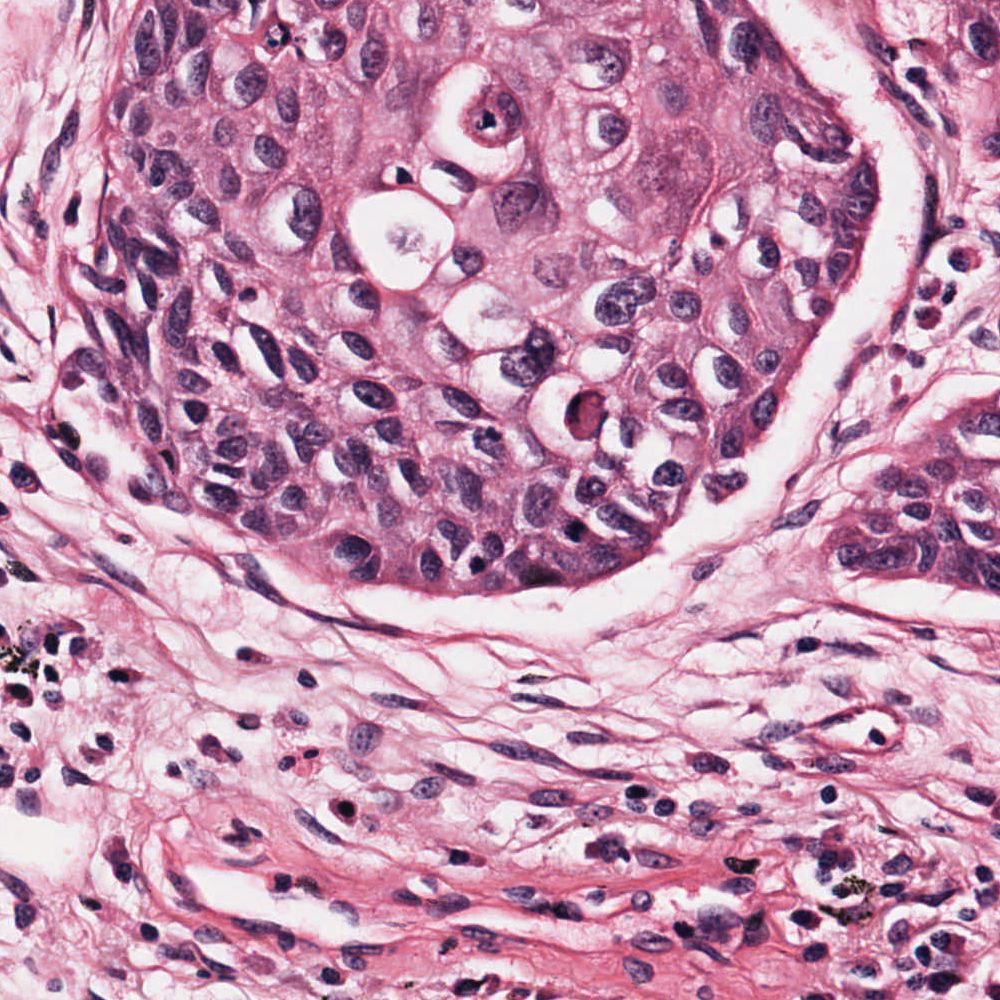} &
		\includegraphics[width=1.6cm]{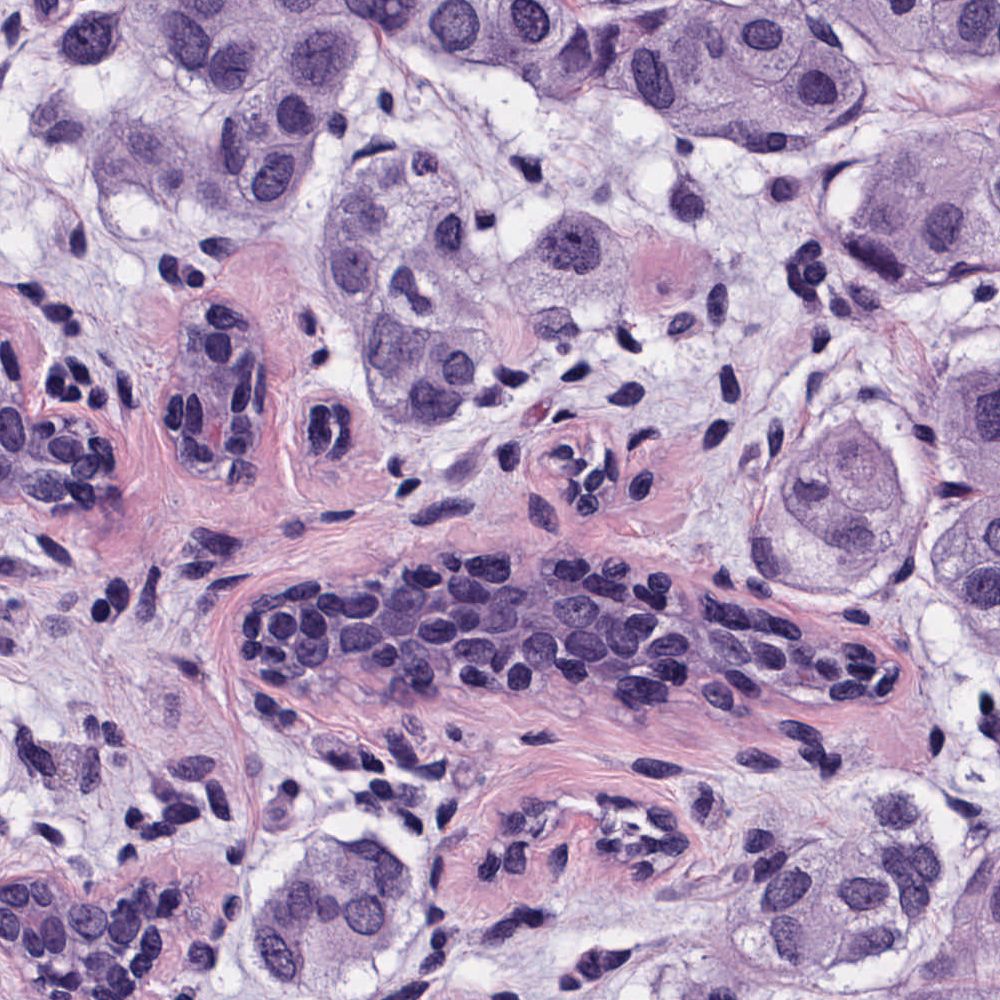} &
		\includegraphics[width=1.6cm]{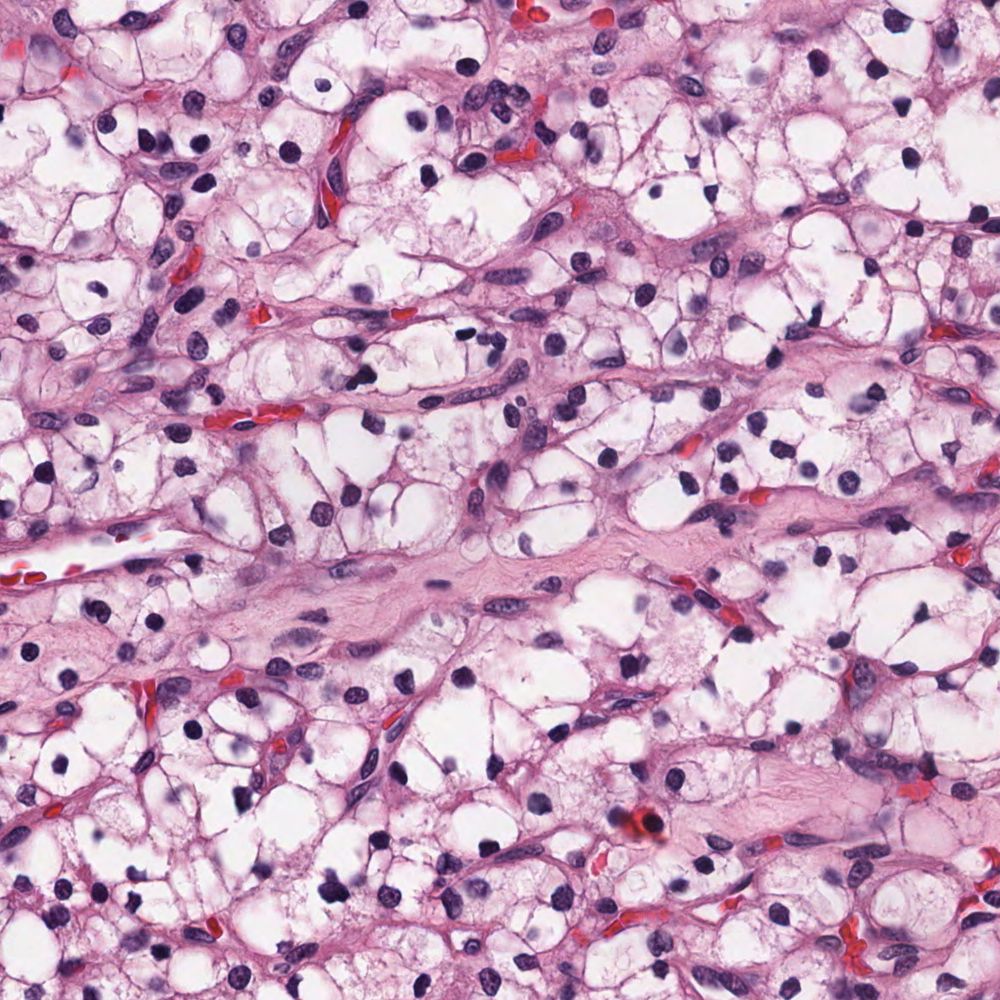} &
		\includegraphics[width=1.6cm]{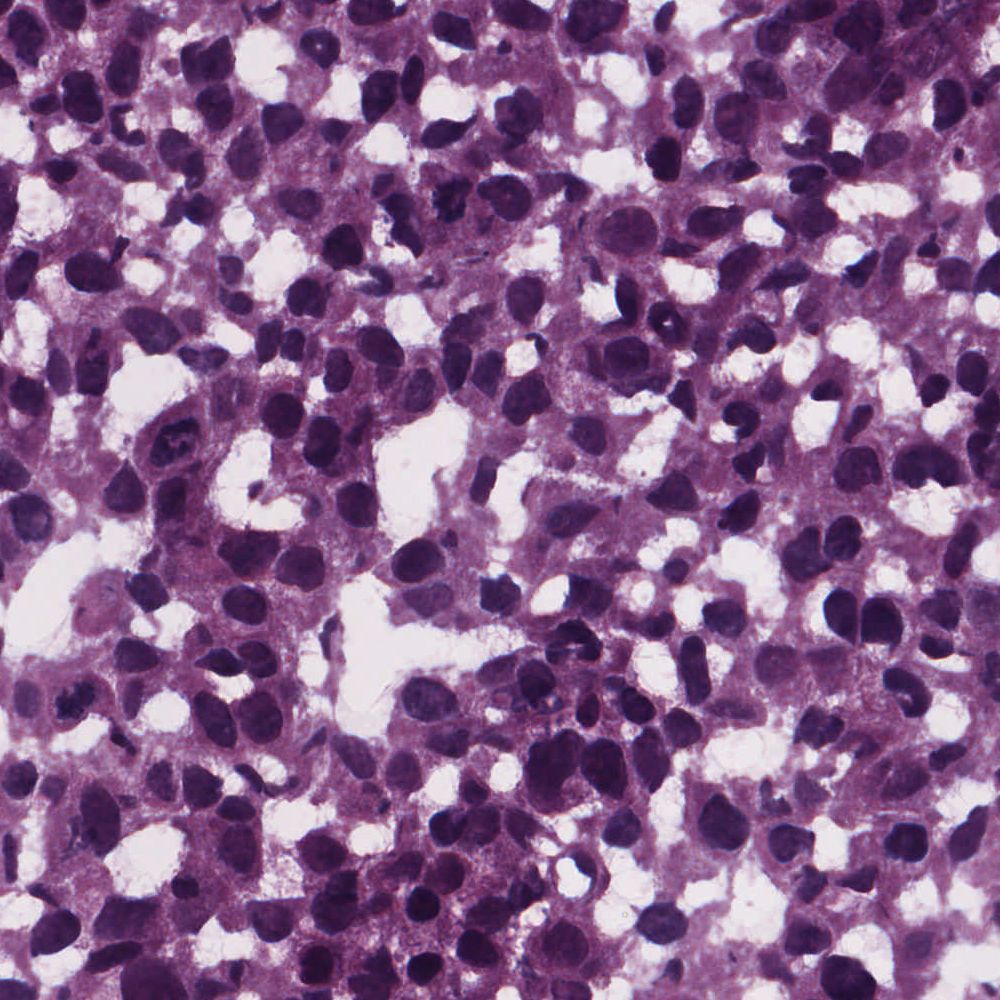} &
		\includegraphics[width=1.6cm]{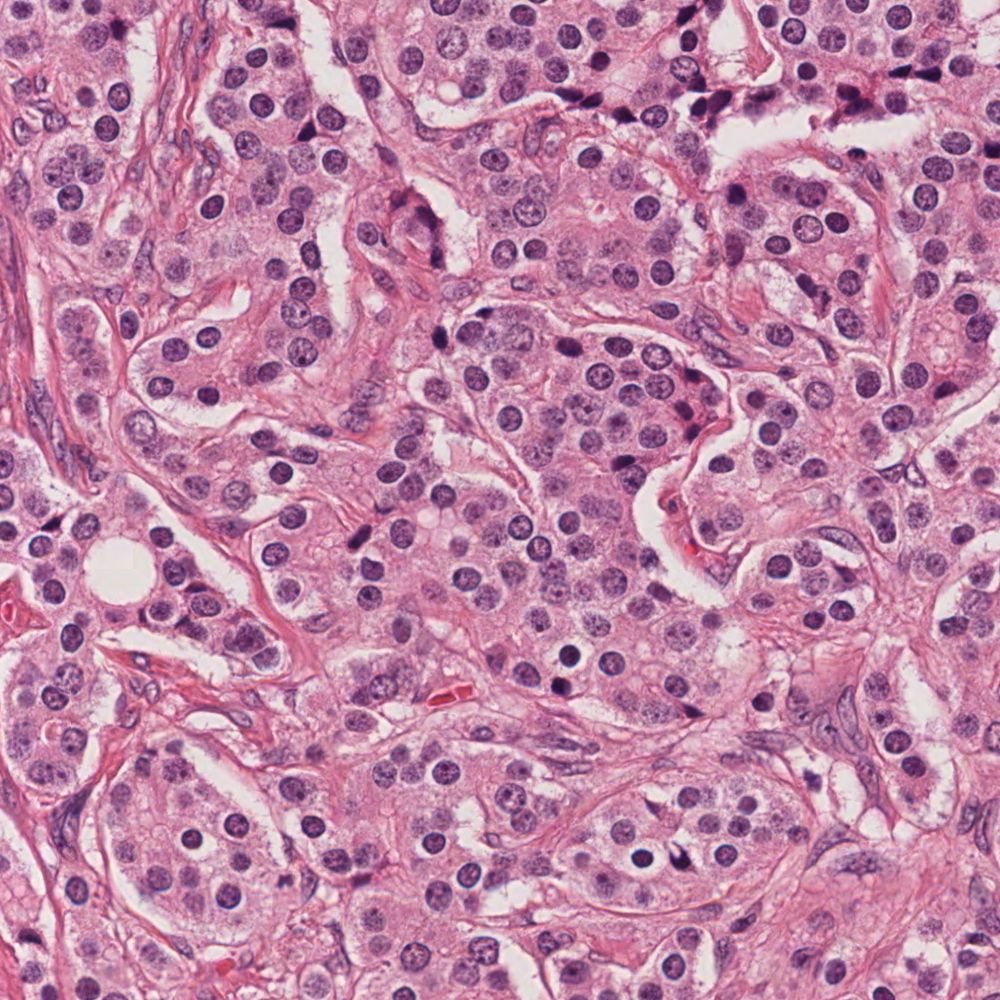} &
		\includegraphics[width=1.6cm]{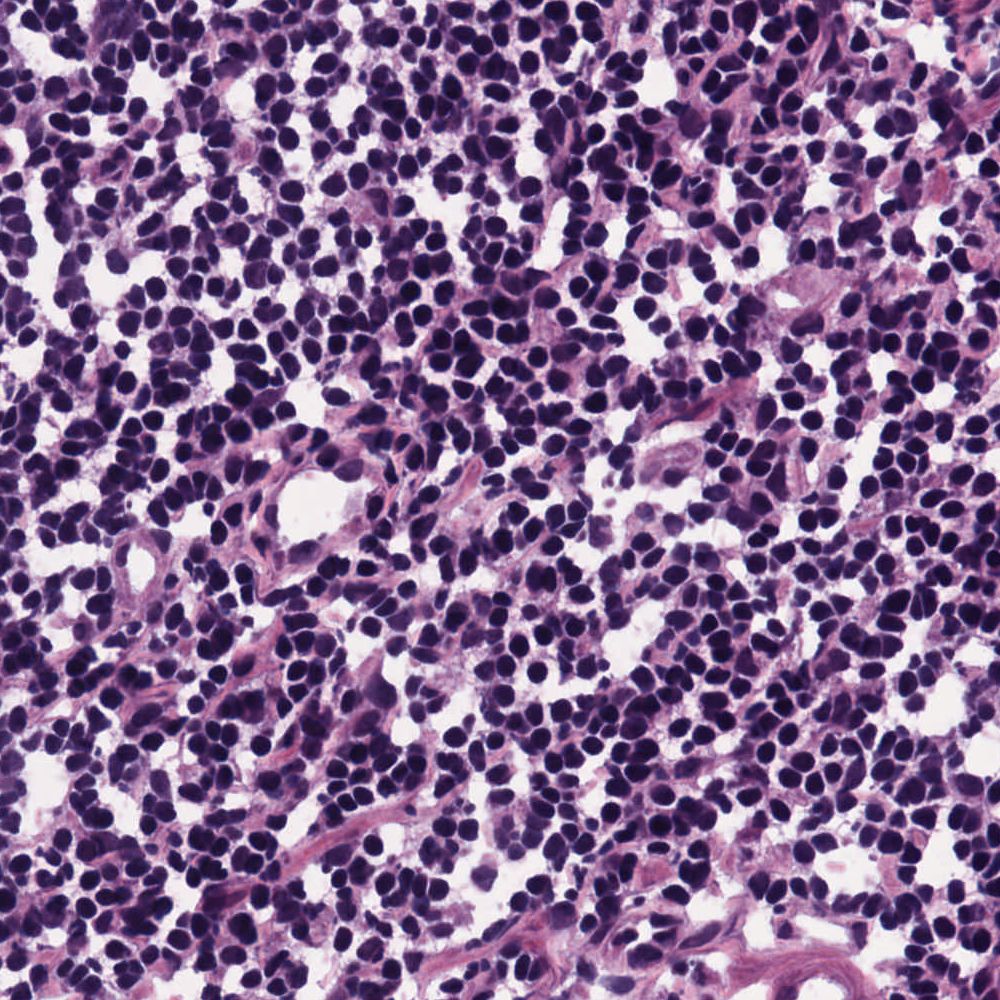} &
		\includegraphics[width=1.6cm]{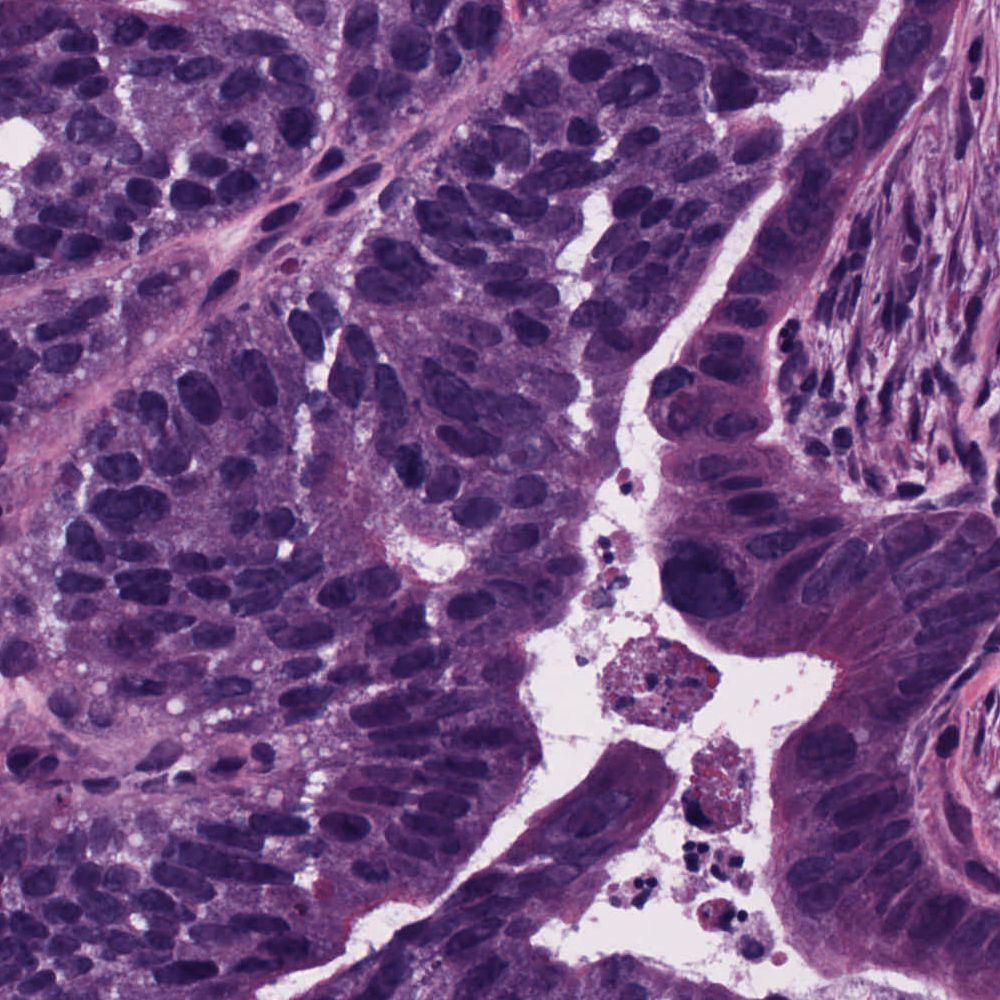}

	\end{tabular}
	\caption{Selected reference images from the MoNuSeg training data based on histogram analysis.}
	\label{all_refs}
\end{figure}

For visualization, examples of selected and non-selected reference images are shown in Fig.~\ref{fig:histogram}. This approach allowed us to automatically select images, such that the colors in the tissue are different from the ones found in the nuclei, ensuring a good contrast of the nuclear boundaries and high visibility of the nuclei. We selected seven reference images (one per organ) using the described technique. The chosen reference images from the MoNuSeg training data are shown in Fig.~\ref{all_refs}.

\begin{figure}
	\centering
	\includegraphics[width=\textwidth]{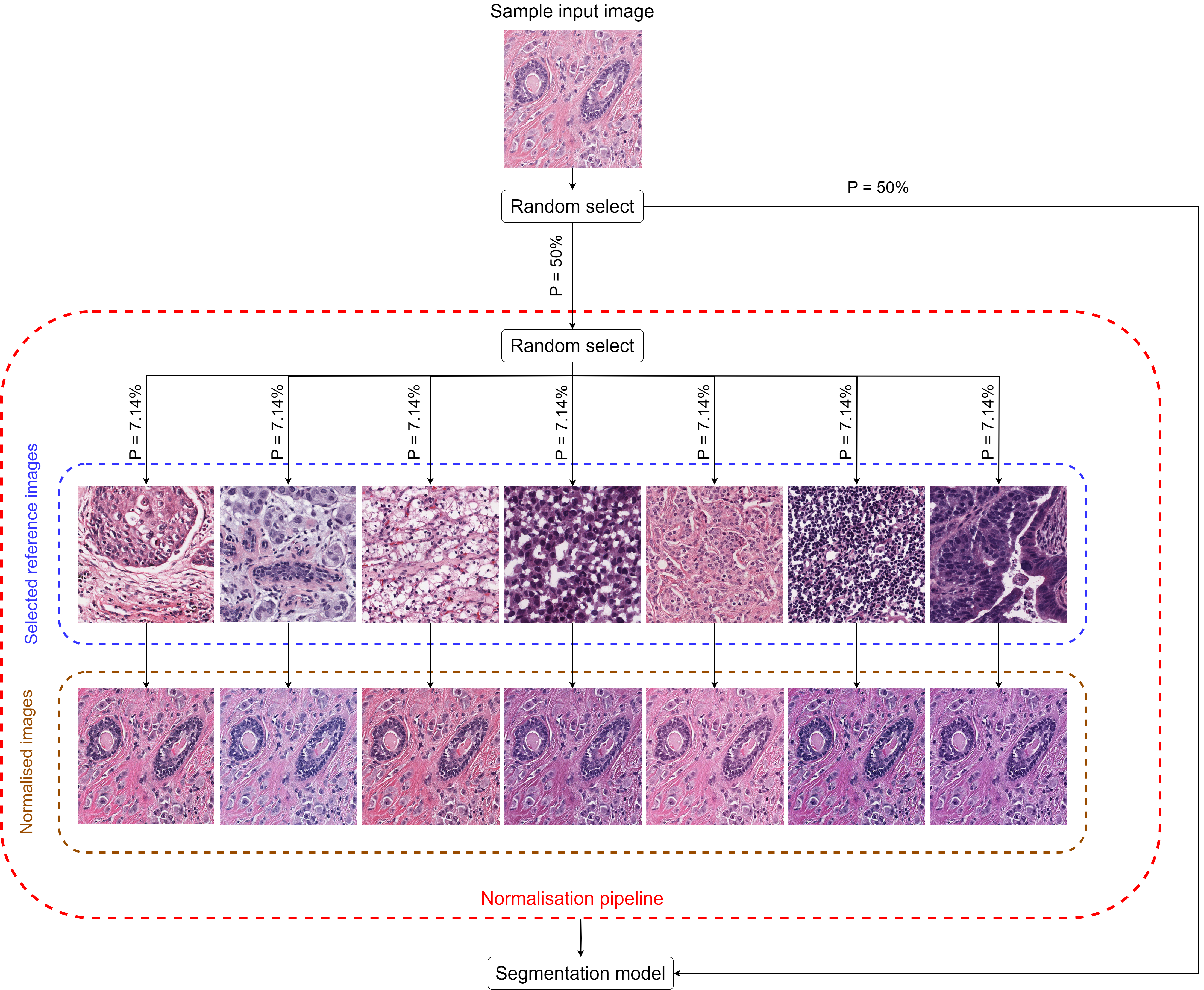}
	\caption{During non-deterministic stain normalization, the input training images were randomly passed either directly to the segmentation model or into the normalization pipeline, where they were normalized to one of seven reference images. The probability for each path in the normalization pipeline was chosen equally.} \label{augmentation}
\end{figure}

We applied non-deterministic stain normalization as an online augmentation based on the selected reference images. For a given input image in the training phase, we either sent the unmodified image to the model (probability of 50\%) or sent it to the normalization pipeline (probability of 50\%). In the latter case, the image was subsequently normalized against one of the reference images based on the Macenko et al. method, introducing another non-deterministic component to the workflow. The probability for each path was chosen equally (i.e., 7.14\% for each path). This step for a sample input image is depicted in Fig.~\ref{augmentation}.

\subsection{Test time stain normalization}
\label{test-time-normalization}
\begin{figure}
	\centering
	\includegraphics[width=\textwidth]{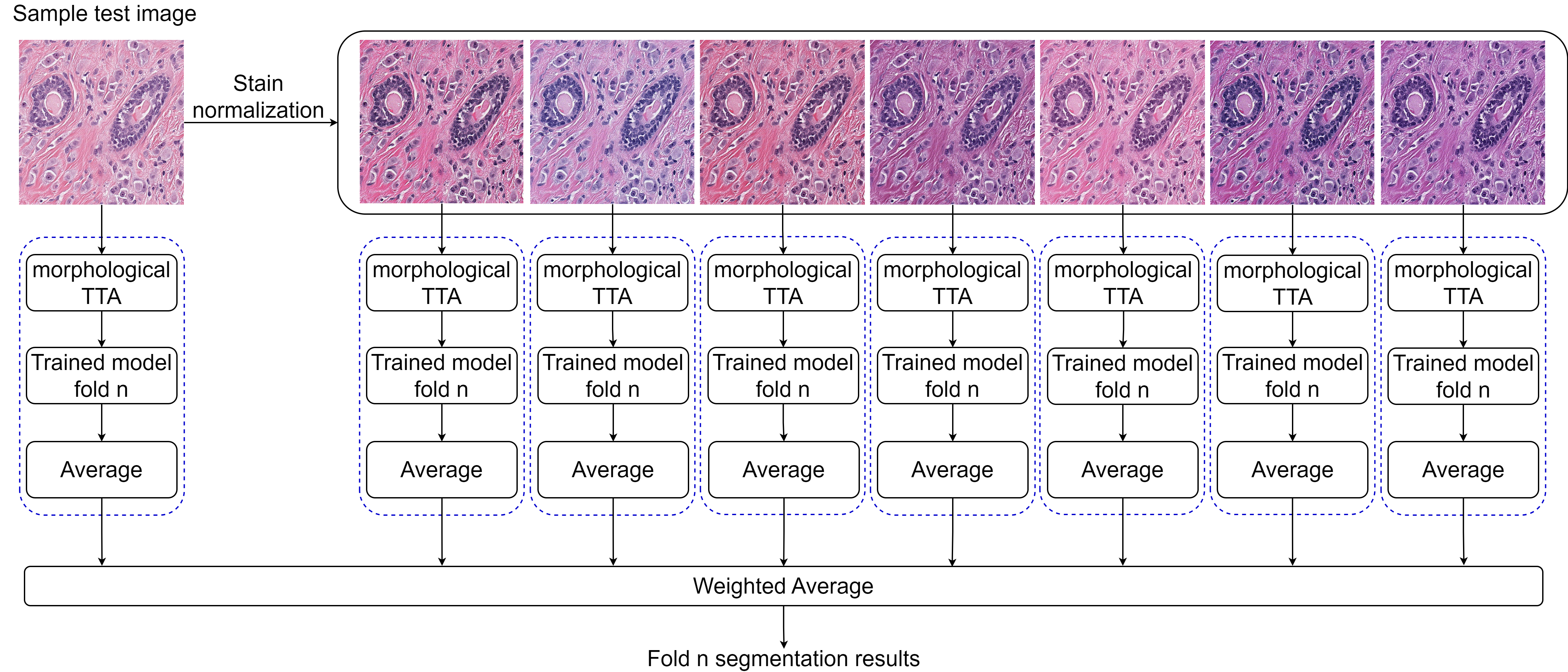}
	\caption{Proposed inference approach with deterministic test time stain normalization. The blue dashed boxes in each branch show the morphological test time augmentation (TTA). Trained model $n\in \{1,2,3,4,5\}$ represents the trained model for each fold of 5-fold cross-validation. } \label{fig:test_normalization}
\end{figure}

Test time augmentation (TTA) has been shown to boost the segmentation and classification performance for various medical image analysis tasks in former studies~\cite{wang2022fuseg, mahbod2021pollen, 9446924}. In this work, besides morphological TTA (90-degree rotation and horizontal flipping), we propose to use test time stain normalization (TTSN). The entire workflow of the inference phase for one of the fold in cross validation (refer to Section~\ref{experimental_setup} for more details) is shown in Fig.~\ref{fig:test_normalization}. In this phase, the original test image and seven normalized test images (identical normalization as performed in training) were sent first to the morphological TTA block (blue boxes in Fig.~\ref{fig:test_normalization}) and then to the trained model. After averaging the results of the morphological TTA block (non-weighted averaging), the outputs were merged using a weighted average scheme. The weights were chosen based on the exploited probabilities in the training phase (i.e., 50 for the original input test image and 7.14 for each of the seven normalized images). 

\subsection{Evaluation}
To evaluate the performance of the nuclei instance segmentation, we used the Dice, aggregate jaccard index (AJI), and panoptic quality (PQ) scores~\cite{Kirillov_2019_CVPR, graham2019hover}. These metrics have been widely used to evaluate and compare the performance of different nuclei segmentation methods~\cite{9745890, monuseg}. While the Dice score shows the general performance of semantic segmentation, AJI and PQ score are sensitive to the capability of the model to separate touching objects (show the instance segmentation performance) and hence are more critical metrics in this study. 

\subsection{Experimental setup}
\label{experimental_setup}
To show the effectiveness and generalization power of the proposed method, we designed six experiments. The schematic workflow of the experiments is shown in Fig.~\ref{fig:experiment}. The details of the experiments are as follows:

\begin{figure}
	\centering
	\includegraphics[width=\textwidth]{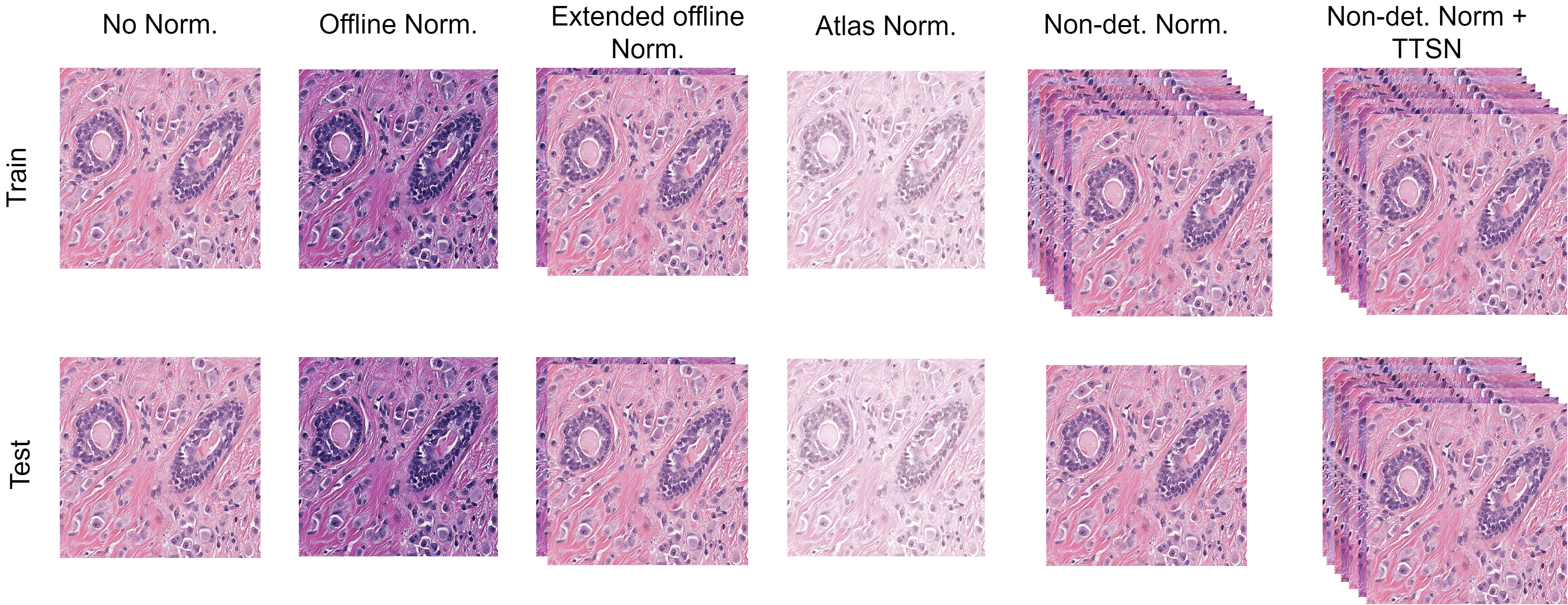}
	\caption{ Schematic design of the experiments. The selected shown image and normalized images are derived from the MoNuSeg training set for visualization. Norm.: Normalization; Non-det.: Non-deterministic; TTSN: Test Time Stain Normalization.} 
	\label{fig:experiment}
\end{figure}

\begin{itemize}
	\item \textbf{No normalization (baseline)}: In this experiment, we did not apply any normalization and just used the raw MoNuSeg training dataset to train the DDU-Net model. To evaluate the model on the test datasets, raw test images were used.   
	\item \textbf{Offline normalization}: In this setup, one single image was chosen as the reference image, and all other images (from the MoNuSeg training set and all test images from all test datasets) were normalized using the Macenko et al. approach. To choose the reference image, we used the histogram analysis described in Section~\ref{Non-deterministic-norm}. 
	\item \textbf{Extended offline normalization}: In this setup,  we merged the raw and normalized MoNuseg training data to train the model (hence, the size of the training set was doubled). In the inference phase, we sent the raw and normalized test images to the trained model and got the average over the results to have the final segmentation mask. Similar to the previous experiment, histogram analysis was used to select the reference image from the MoNuSeg training dataset. 
	\item \textbf{Offline normalization with atlas image}: In this experiment, we used an atlas image to perform offline normalization for all training images. The atlas image was formed by utilizing the average of all seven selected reference images as described in Section~\ref{Non-deterministic-norm}. The average image was then used to perform Macenko et al. normalization. In the inference phase, the same atlas image was used to perform offline stain normalization for all test images in all test datasets.
	
	\item \textbf{Non-deterministic stain normalization}: In this experiment, we applied the proposed non-deterministic stain normalization as described in Section~\ref{Non-deterministic-norm}. In the inference phase, the raw test images were sent to the trained model. 
	\item \textbf{Non-deterministic normalization + TTSN}: In this experiment, we used both non-deterministic stain normalization and TTSN as described in Section~\ref{Non-deterministic-norm} and Section~\ref{test-time-normalization}. 
\end{itemize}

Besides the differences mentioned above, all other parameters were kept identical in all experiments. We used 5-fold cross-validation to train the DDU-Net in all experiments and generated five models from cross-validation ensembled in the inference phase. Each model was trained for 200 epochs with an initial learning rate (LR) of 0.001. We used a LR scheduler (dropped by half after every 30 epochs). Random cropped images with the size of $992 \times 992$ pixels were used to train the models in the experiments. In the inference phases, all test datasets except CPM-15, test images were white-padded to form $1024 \times 1024$ images and then sent to the train models. The original part was then cropped from the results. For the CPM-15 dataset, all images were white-padded to $1056 \times 1056$ pixels, as one of the images in the CPM-15 dataset has a larger size than $1024$ in one dimension ($1032 \times 808$ pixels). Moreover, Adam optimizer~\cite{Kingma2014}, batch size of four, and a threshold of 0.5 (to convert probabilities to binary values) were used in training and testing. We also applied classical augmentation techniques in all experiments mentioned above.  The classical augmentations included horizontal and vertical flipping, random 90-degree rotations, and random color, brightness and contrast shifts.

To investigate the robustness, the entire experiments (5-fold cross-validation training and ensemble for each of the six setups) were repeated three times, and the average results and standard deviations were reported. All models were trained using a single workstation with an Intel Core i7-8700 3.20 GHz CPU, 32 GB of RAM, and a TITIAN V Nvidia GPU card with 12 GB of installed memory. The DDU-Net was trained and evaluated using Tensorflow (version 2.11) DL framework.

\section{Results and Discussion}
In this section, we report the results for each test dataset in a separate table. Each table contains the results based on the different experimental setups explained in Section~\ref{experimental_setup} including baseline segmentation results (first row), results from offline normalization technique (second row), results from extended offline normalization (third row), results from the offline normalization approach with atlas image (fourth row) and the results from the proposed method with or without TTSN (fifth and sixth rows).

The results in Table~\ref{monuseg_res} are derived from the MoNuSeg test data. Although the test data and train data came from the same distribution (MoNuSeg dataset) in this experiment, the average results indicate a superior segmentation performance of the proposed method (with or without TTSN) in comparison to the baseline results. The proposed method with TTSN also delivered better semantic and instance segmentation performance compared to offline normalization (single image or atlas image) or extended offline normalization approaches. Comparing the fifth and sixth rows shows that using TTSN slightly improves the segmentation performance for all evaluation indices.

The results of Table~\ref{tnbc_res} to Table~\ref{nuinsseg_res} show the generalization capability of the models in different experimental setups for unseen test datasets. A number of observations can be inferred from the results of these tables. 

First of all, for all test datasets, the proposed method (with or without TTSN) consistently delivers better average segmentation performance in comparison to the baseline model for all evaluation metrics. 

Secondly, using conventional stain offline normalization is not always beneficial, and it can even degrade performance (for example, in degrading the average performance for the TNBC dataset (Table~\ref{tnbc_res}) across all evaluation metrics by comparing the first and second rows). This is in accordance with previous findings in other studies~\cite{tellez2019quantifying, aresta2019bach}. However, in this study, we developed an approach that consistently delivers superior performance on multiple test datasets compared to the baseline model.

Thirdly, adding TTSN to the workflow (i.e., comparison between the fifth and sixth rows in the tables) improves the average segmentation performance in most test datasets (5 out 7 based on Dice score, 5 out of 7 for AJI, 5 out of 7 for PQ score) and delivers very competitive performance in other cases. However, it is worth mentioning that adding the proposed TTSN to the workflow would increase the test time by eight folds (instead of sending one single test image to the models, eight images have to be sent to the model as shown in Fig.~\ref{fig:test_normalization}). This could be a barrier of using TTSN when limited computational resources are available to analyze whole slide histological images. However, with proper computational resources and parallel processing, this should not be an issue. To investigate if incorporating more images in the ensemble phase would change the segmentation results, we performed an additional experiment with 14 images for deterministic stain normalization. We chose two images per organ instead of choosing one image per organ. Again, we used the absolute differences between the mean tissue and nuclear intensity to select the second image per organ (the second suitable candidate for each organ). The results are reported in Table~\ref{images14}. As the results indicate, the performance is almost identical in all cases (We reported the results with two decimal precision in the table as they are very close). This suggests that adding more images for deterministic stain normalization in the inference phase does not necessarily lead to improved performance but increases the inference time, which is undesirable for practical applications.

Fourthly, for most cases (all except the CryoNuSeg dataset), the proposed method with TTSN delivers superior average semantic and instance segmentation performance compared to other normalization methods discussed in this study (offline normalization (single image or atlas image) and extended offline normalization).

\begin{figure}[t!]
	\centering
	\begin{tabular}{cccc}
		Original image & Ground truth & Base prediction & Proposed prediction\\
		\includegraphics[width=3cm]{figs/overlay/human_bladder_12.png} &
		\includegraphics[width=3cm]{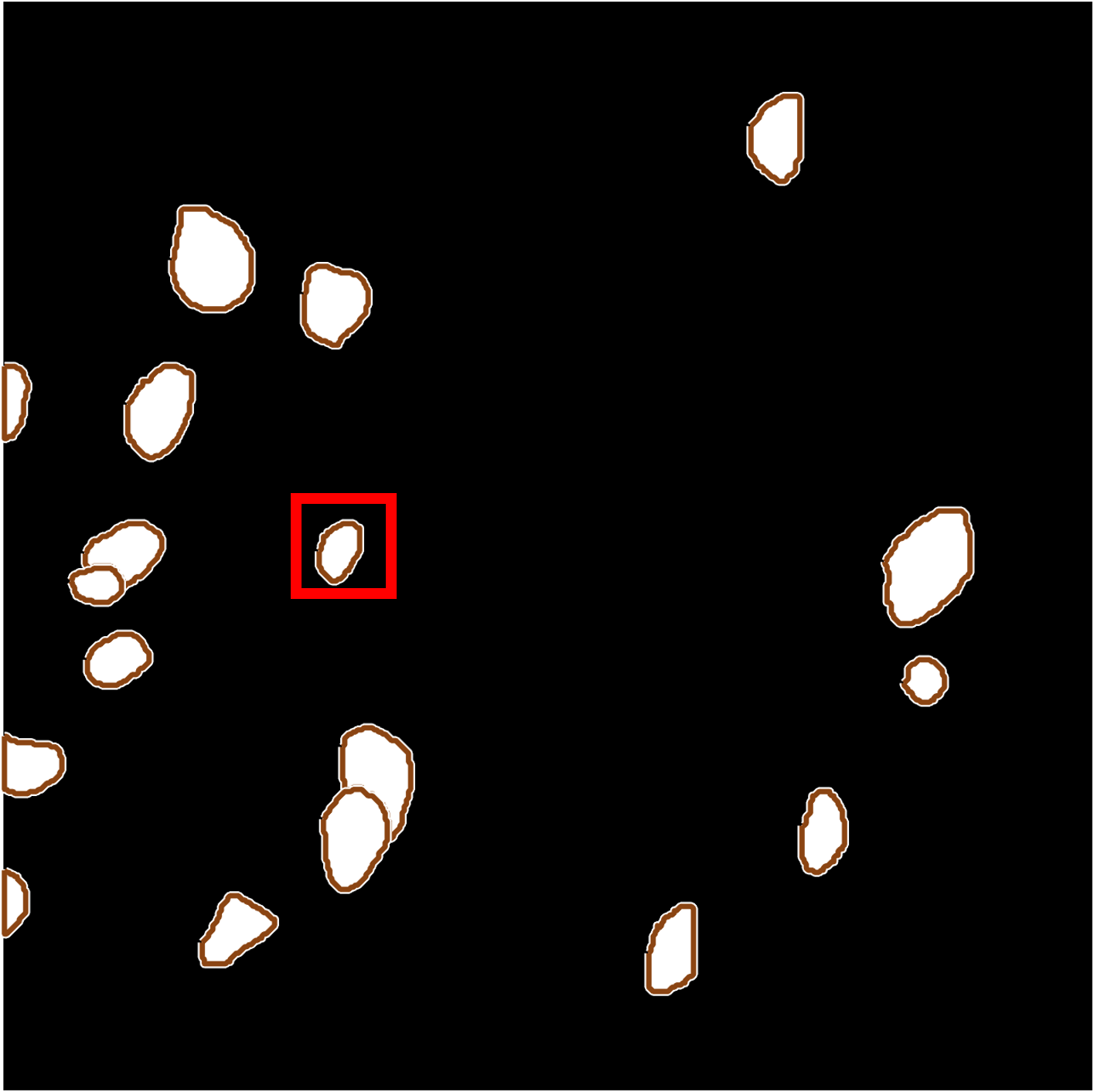} &
		\includegraphics[width=3cm]{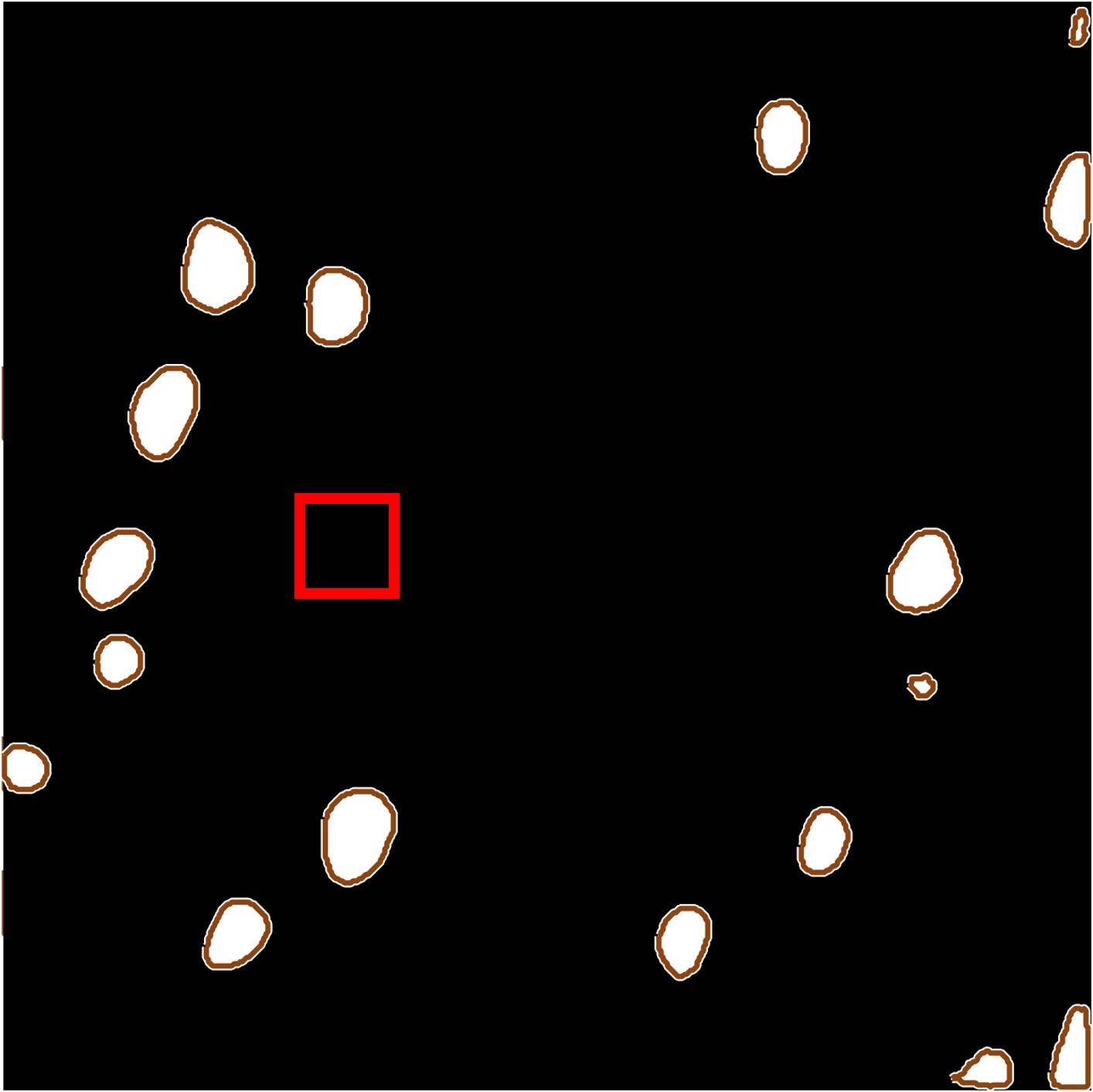} &
		\includegraphics[width=3cm]{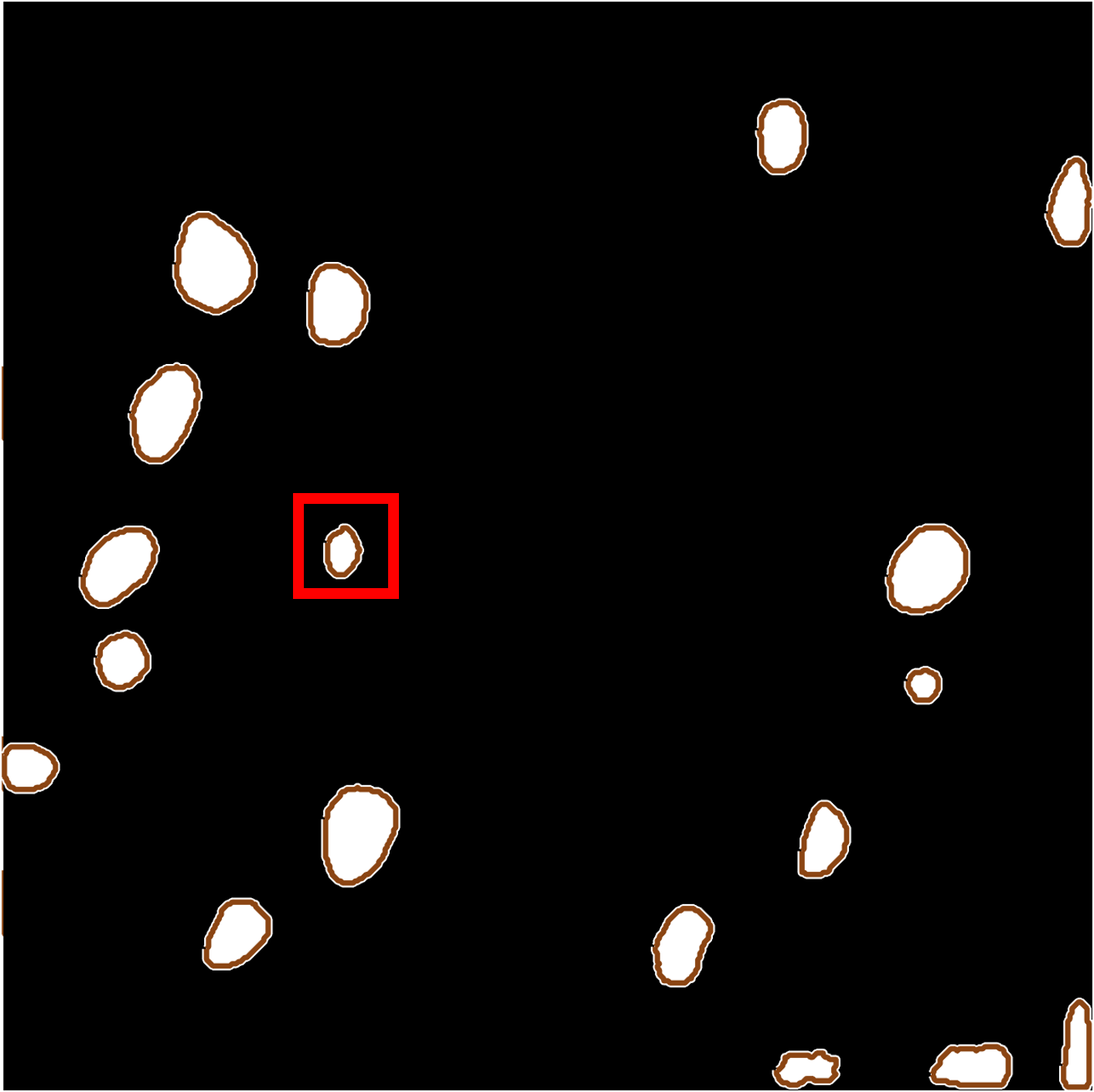} 
		\\
		\includegraphics[width=3cm]{figs/overlay/human_placenta_07.png} &
		\includegraphics[width=3cm]{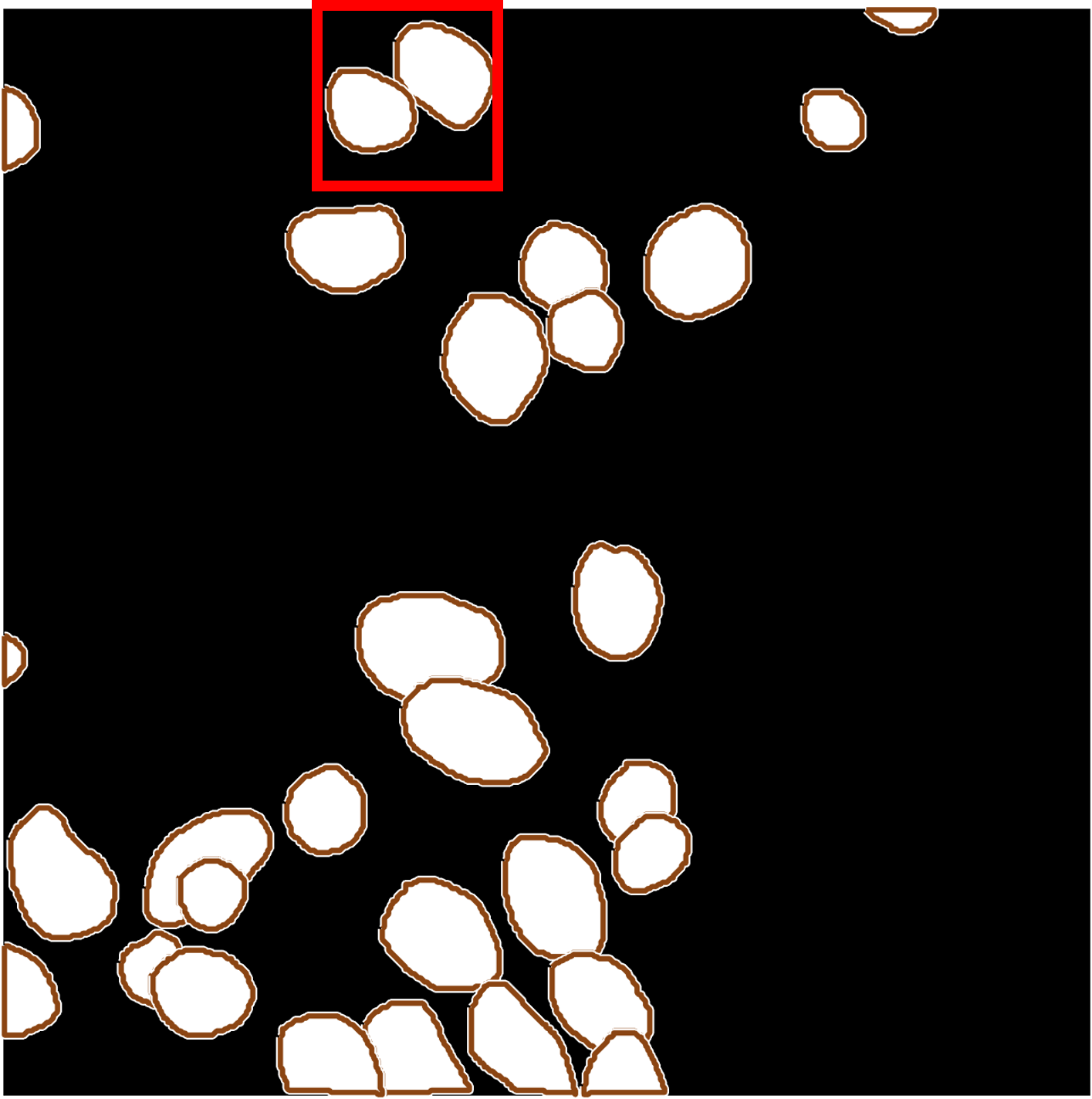} &
		\includegraphics[width=3cm]{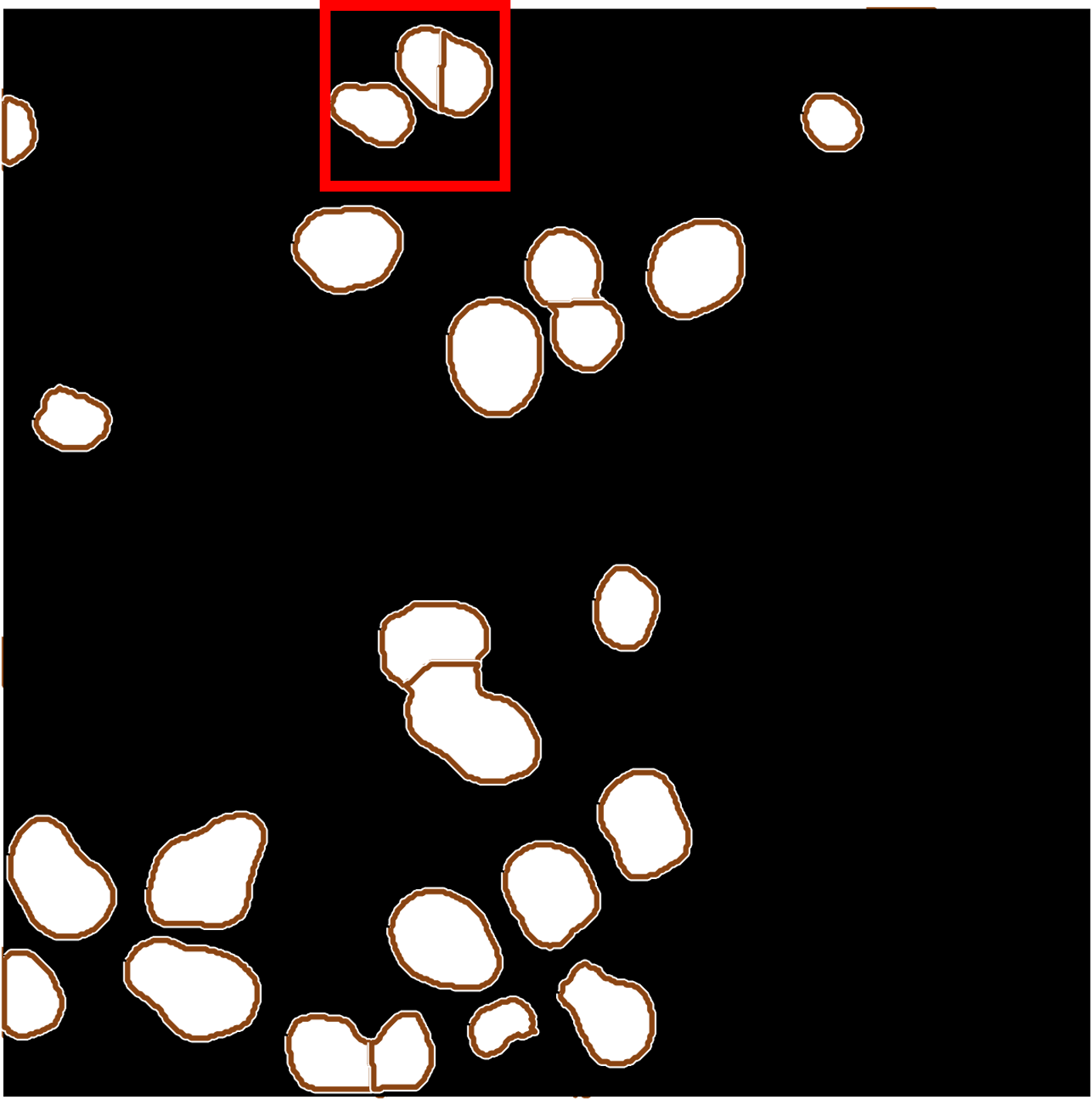} &
		\includegraphics[width=3cm]{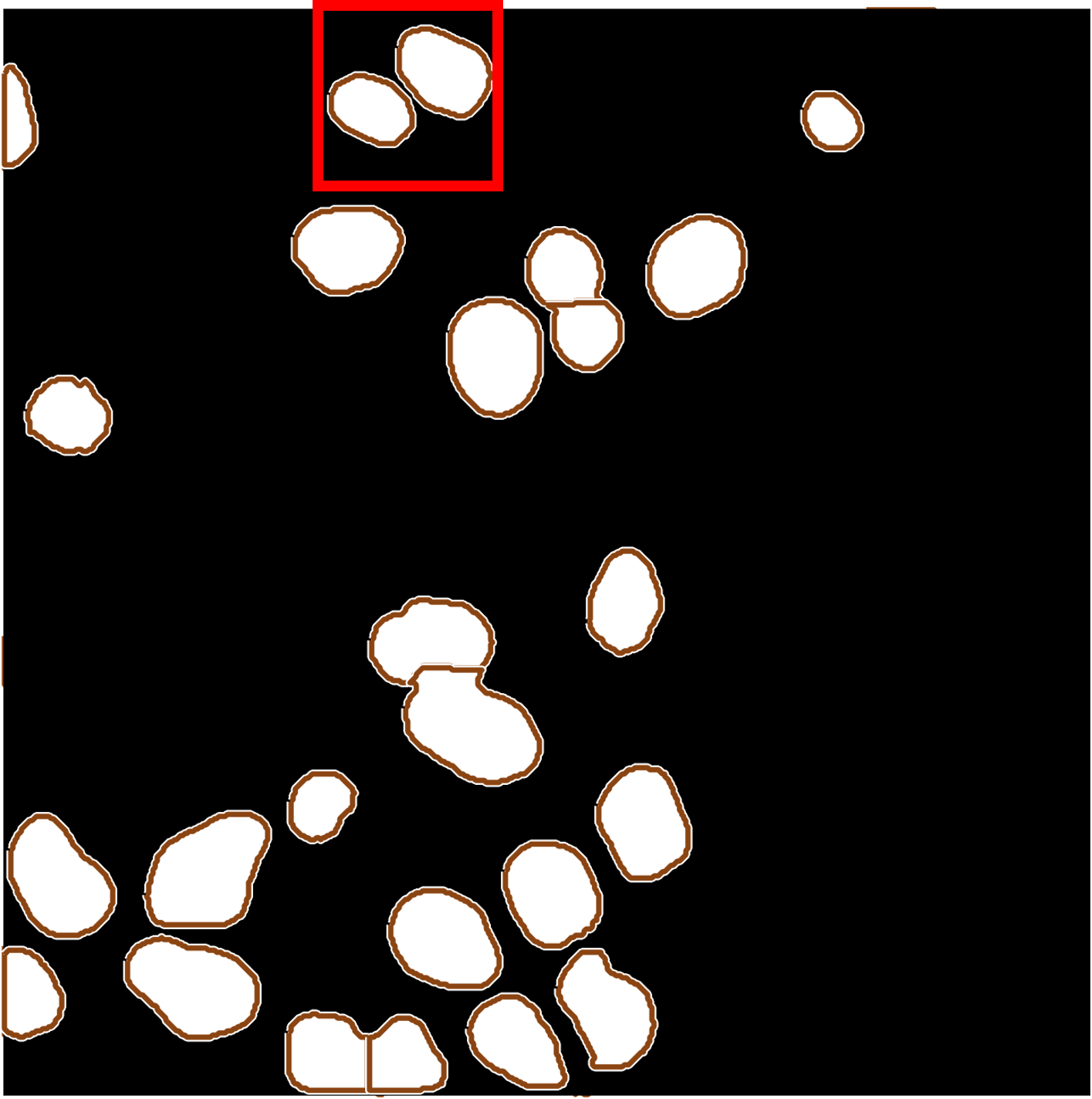} 
		\\
		\includegraphics[width=3cm]{figs/overlay/human_cerebellum_2.png} &
		\includegraphics[width=3cm]{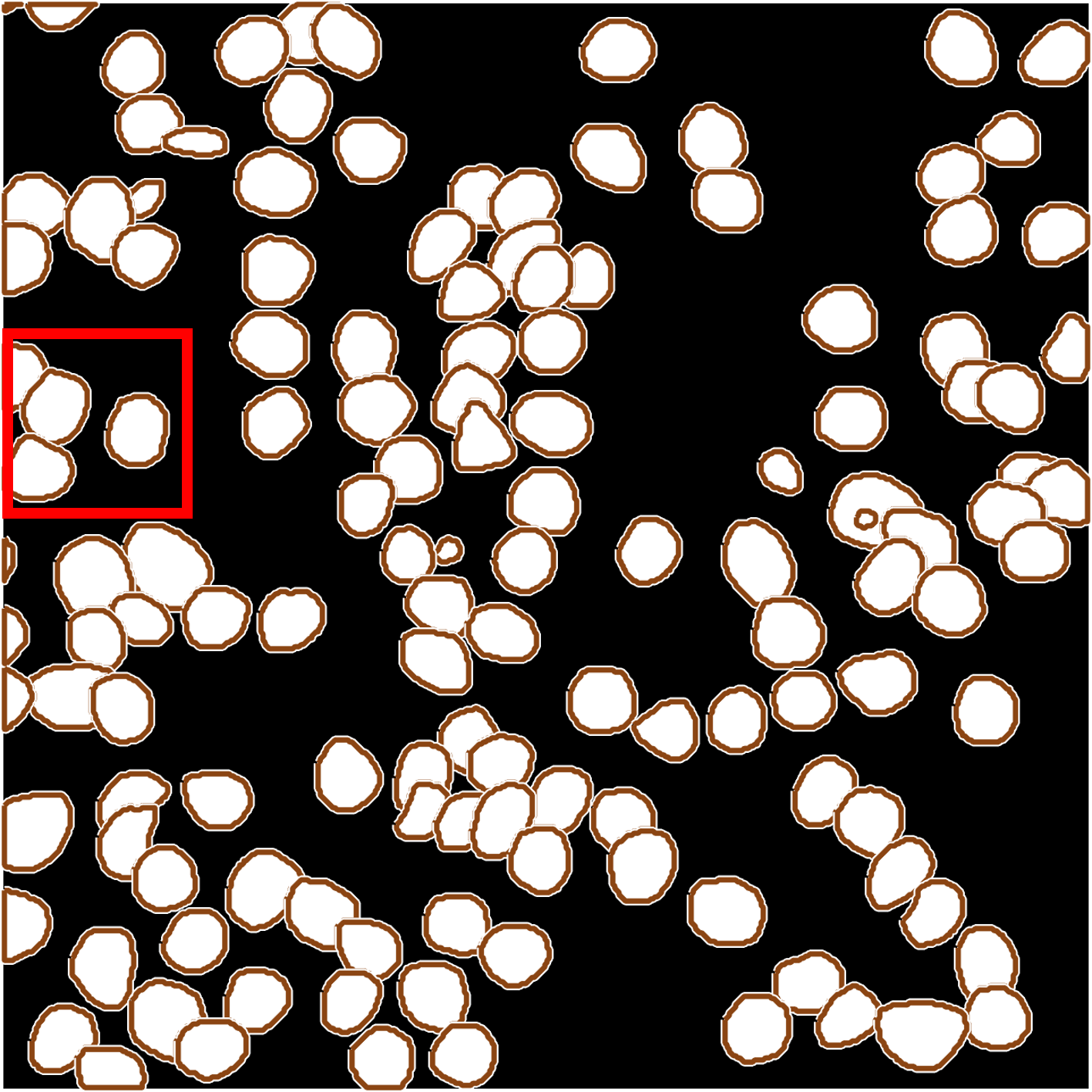} &
		\includegraphics[width=3cm]{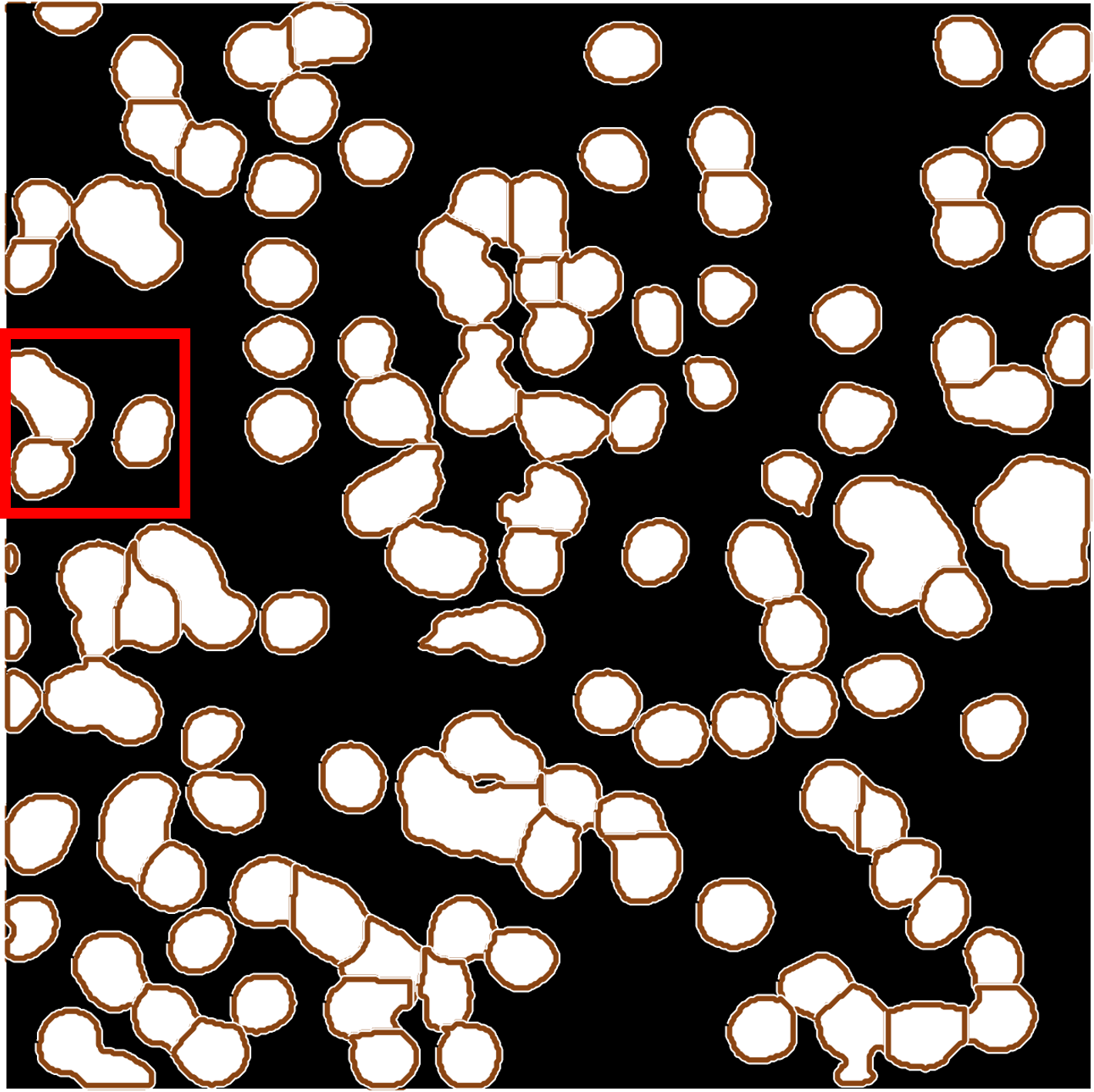} &
		\includegraphics[width=3cm]{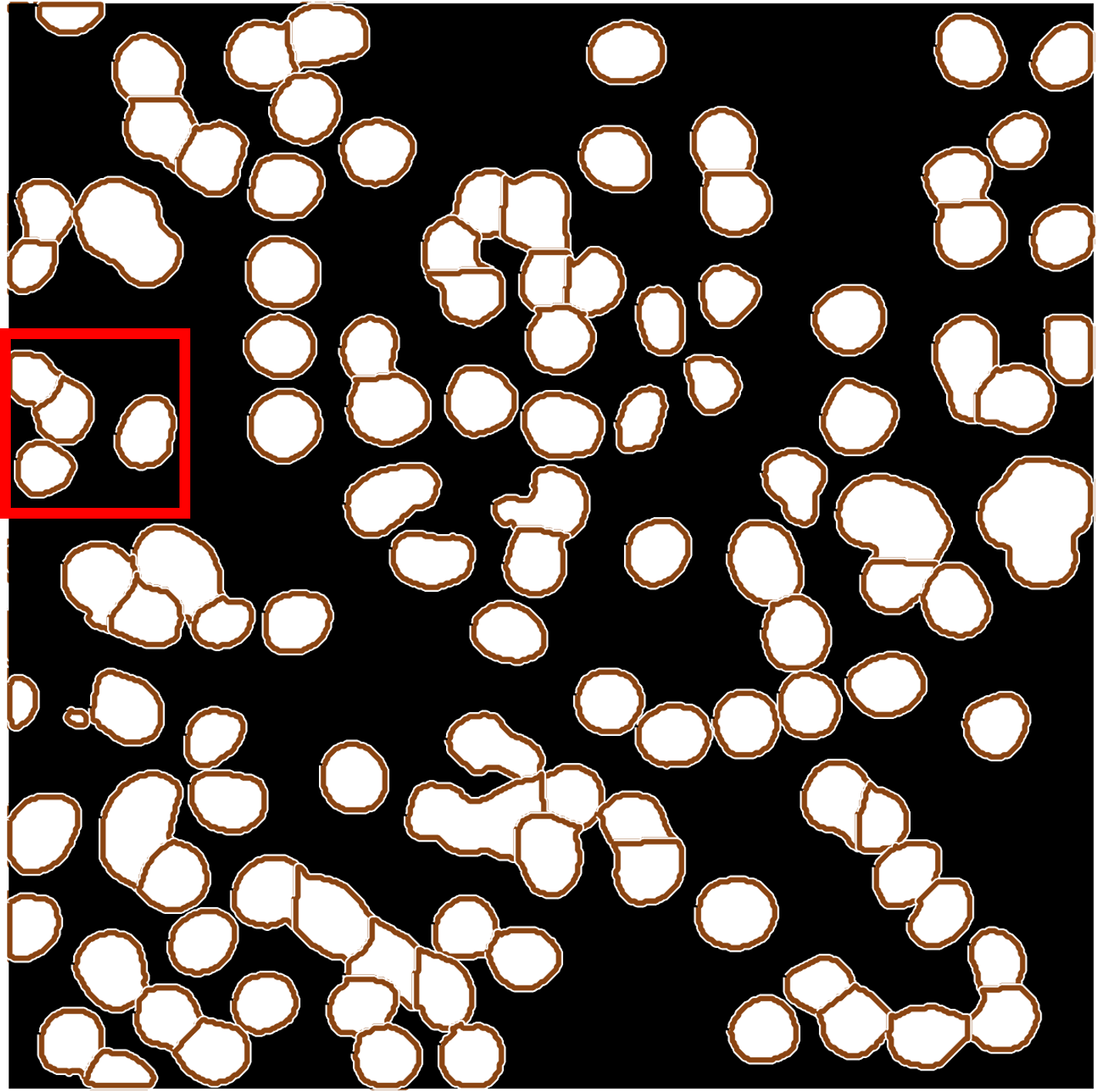} 
		
	\end{tabular}
	\caption{Qualitative comparison between baseline segmentation model and proposed approach (results from the first run). The first column shows some example test images from the NuInsSeg dataset (from the human bladder, human placenta, and human cerebellum in the first, second, and third rows, respectively). The second column shows the ground truth segmentation masks. The third column shows the prediction by the baseline DDU-Net model, and the fourth column shows the results of the proposed approach. The red bounding boxes in columns two to four show some example nuclei where the proposed method delivered a superior segmentation performance (better semantic segmentation performance in the first row and better instance segmentation performance in the second and third rows) compared to the baseline segmentation model.}
	\label{fig:compare_res}
\end{figure}

\begin{table}
	\centering
	\caption{Segmentation results for the MoNuSeg test data (average and standard deviation over three runs). The last two rows represent the results from the proposed approach. Non-det.: Non-deterministic; TTSN: Test Time Stain Normalization; AJI: Aggregate Jaccard Index; PQ: Panoptic Quality.}\label{monuseg_res}
	\begin{tabular}{l|c|c|c}
		\hline
		Method &  Dice (\%) & \ AJI (\%) & PQ (\%)   \\
		\hline
		No normalization (baseline)                    & 82.6 $\pm$ 0.2 & 65.5 $\pm$ 0.2  & 63.5 $\pm$ 0.1 \\
		Offline normalization                          & 83.1 $\pm$ 0.1 & 66.1 $\pm$ 0.1  & 63.5 $\pm$ 0.2 \\
		Extended offline normalization                 & 83.1 $\pm$ 0.1 & 66.8 $\pm$ 0.1  & 64.6 $\pm$  0.3  \\
		Offline normalization with atlas               & 83.2 $\pm$ 0.1 & 65.8 $\pm$ 0.2  & 63.3 $\pm$ 0.3\\\hline
		Non-det. stain normalization                   & 83.0 $\pm$ 0.1 & 66.6 $\pm$ 0.1  & 64.5 $\pm$ 0.2   \\
		Non-det. stain normalization + TTSN            & 83.4 $\pm$ 0.1 & 66.9 $\pm$ 0.2  & 64.9 $\pm$ 0.3 \\

		\hline
	\end{tabular}
\end{table}
\begin{table}
	\centering
	\caption{Segmentation results for the TNBC test data (average and standard deviation over three runs). The last two rows represent the results from the proposed approach. Non-det.: Non-deterministic; TTSN: Test Time Stain Normalization; AJI: Aggregate Jaccard Index; PQ: Panoptic Quality.}\label{tnbc_res}
	\begin{tabular}{l|c|c|c}
		\hline
		Method &  Dice (\%) & \ AJI (\%) & PQ (\%)   \\
		\hline
		No normalization (baseline)                    & 79.6 $\pm$ 1.3 & 61.4 $\pm$ 1.5  &  59.3 $\pm$ 1.3 \\
		Offline normalization                          & 78.4 $\pm$ 0.8 & 60.8 $\pm$ 0.7  &  58.1 $\pm$ 0.5\\
		Extended offline normalization                 & 78.9 $\pm$ 0.6 & 61.4 $\pm$ 0.7  &  60.6 $\pm$ 0.4   \\
		Offline normalization with atlas               & 79.1 $\pm$ 0.6 & 61.5 $\pm$ 0.6  &  59.2 $\pm$ 0.2\\\hline
		Non-det. stain normalization                   & 81.3 $\pm$ 0.2 & 64.6 $\pm$ 0.3  &  62.9 $\pm$ 0.5  \\
		Non-det. stain normalization + TTSN            & 81.4 $\pm$ 0.2 & 64.8 $\pm$ 0.2  &  62.3 $\pm$ 0.4\\
		
		\hline
	\end{tabular}
\end{table}
\begin{table}
	\centering
	\caption{Segmentation results for the CryoNuSeg test data (average and standard deviation over three runs). The last two rows represent the results from the proposed approach. Non-det.: Non-deterministic; TTSN: Test Time Stain Normalization; AJI: Aggregate Jaccard Index; PQ: Panoptic Quality.}\label{cryonuseg_res}
	\begin{tabular}{l|c|c|c}
		\hline
		Method &  Dice (\%) & \ AJI (\%) & PQ (\%)   \\
		\hline
		No normalization (baseline)                    & 78.4 $\pm$ 0.2 & 50.6 $\pm$ 0.1  & 47.3 $\pm$ 0.1  \\
		Offline normalization                          & 78.9 $\pm$ 0.9 & 50.4 $\pm$ 0.5  & 47.1 $\pm$ 0.4\\
		Extended offline normalization                 & 79.3 $\pm$ 0.2 & 52.3 $\pm$ 0.2  & 49.2 $\pm$ 0.2  \\
		Offline normalization with atlas               & 78.4 $\pm$ 0.1 & 50.5 $\pm$ 0.4  & 47.2 $\pm$ 0.2 \\\hline
		Non-det. stain normalization                   & 79.1 $\pm$ 0.2 & 52.1 $\pm$ 0.2  & 49.1 $\pm$ 0.2    \\
	    Non-det. stain normalization + TTSN            & 79.0 $\pm$ 0.2 & 52.0 $\pm$ 0.1  & 49.2 $\pm$ 0.2\\
		\hline
	\end{tabular}
\end{table}
\begin{table}
	\centering
	\caption{Segmentation results for the CPM-15 test data (average and standard deviation over three runs). The last two rows represent the results from the proposed approach. Non-det.: Non-deterministic; TTSN: Test Time Stain Normalization; AJI: Aggregate Jaccard Index; PQ: Panoptic Quality.}\label{cpm15_res}
	\begin{tabular}{l|c|c|c}
		\hline
		Method &  Dice (\%) & \ AJI (\%) & PQ (\%)   \\
		\hline
		No normalization (baseline)                    & 74.0 $\pm$ 0.3 & 53.3 $\pm$ 0.2 & 49.6 $\pm$ 0.2 \\
		Offline normalization                          & 74.4 $\pm$ 0.6 & 53.7 $\pm$ 0.7 & 49.4 $\pm$ 0.8\\
	    Extended offline normalization                 & 73.3 $\pm$ 0.3 & 52.9 $\pm$ 0.1 & 48.9 $\pm$ 0.1   \\
	    Offline normalization with atlas               & 75.2 $\pm$ 0.2 & 54.2 $\pm$ 0.2 & 50.3 $\pm$ 0.3\\\hline
		Non-det. stain normalization                   & 75.3 $\pm$ 0.5 & 54.3 $\pm$ 0.6 & 50.4 $\pm$ 0.9     \\
		Non-det. stain normalization + TTSN            & 75.7 $\pm$ 0.6 & 54.6 $\pm$ 0.6 & 50.8 $\pm$ 0.9\\
		
		\hline
	\end{tabular}
\end{table}
\begin{table}
	\centering
	\caption{Segmentation results for the CPM-17 test data (average and standard deviation over three runs). The last two rows represent the results from the proposed approach. Non-det.: Non-deterministic; TTSN: Test Time Stain Normalization; AJI: Aggregate Jaccard Index; PQ: Panoptic Quality.}\label{cpm17_res}
	\begin{tabular}{l|c|c|c}
		\hline
		Method &  Dice (\%) & \ AJI (\%) & PQ (\%)   \\
		\hline
		No normalization (baseline)                    & 81.2 $\pm$ 0.1 & 62.6 $\pm$ 0.2  & 58.3 $\pm$ 0.4 \\
		Offline normalization                          & 82.0 $\pm$ 0.2 & 63.8 $\pm$ 0.4  & 58.9 $\pm$ 0.5 \\
		Extended offline normalization                 & 80.7 $\pm$ 0.4 & 62.6 $\pm$ 0.7  & 58.4 $\pm$ 0.7   \\
		Offline normalization with atlas               & 80.2 $\pm$ 0.2 & 63.8 $\pm$ 0.3  & 58.9 $\pm$ 0.2\\\hline
		Non-det. stain normalization                   & 81.9 $\pm$ 0.3 & 63.9 $\pm$ 0.6  & 59.6 $\pm$ 0.9 \\
		Non-det. stain normalization + TTSN            & 82.3 $\pm$ 0.4 & 64.3 $\pm$ 0.6  & 59.7 $\pm$ 0.8\\
		
		\hline
	\end{tabular}
\end{table}
\begin{table}
	\centering
	\caption{Segmentation results for the ConSep test data (average and standard deviation over three runs). The last two rows represent the results from the proposed approach. Non-det.: Non-deterministic; TTSN: Test Time Stain Normalization; AJI: Aggregate Jaccard Index; PQ: Panoptic Quality.}\label{consep_res}
	\begin{tabular}{l|c|c|c}
		\hline
		Method &  Dice (\%) & \ AJI (\%) & PQ (\%)   \\
		\hline
		No normalization (baseline)                    & 70.1 $\pm$ 0.8 & 40.6 $\pm$ 0.2 & 39.7 $\pm$ 0.2 \\
		Offline normalization                          & 66.7 $\pm$ 0.7 & 36.8 $\pm$ 0.9 & 35.1 $\pm$ 0.8 \\
		Extended offline normalization                 & 68.7 $\pm$ 1.5 & 41.5 $\pm$ 1.1 & 41.2 $\pm$ 1.3  \\
		Offline normalization with atlas               & 68.3 $\pm$ 0.6 & 38.8 $\pm$ 0.6 & 37.5$\pm$ 0.3\\\hline
		Non-det. stain normalization                   & 72.4 $\pm$ 0.7 & 43.5 $\pm$ 0.8 & 43.1 $\pm$ 0.8  \\
		Non-det. stain normalization + TTSN            & 71.5 $\pm$ 0.1 & 42.7 $\pm$ 0.3 & 41.9 $\pm$ 0.1 \\
		
		\hline
	\end{tabular}
\end{table}
The results in Table~\ref{nuinsseg_res} show the segmentation performance for the NuInsSeg dataset. NuInSeg is the largest test dataset used in this study, with 665 images that are derived from 31 different human and mouse organs. Therefore, this dataset can be considered the most important dataset to show the generalization capability of the proposed method, and hence, we describe its results separately. As the results show, we observe the same trend of superior performance of the proposed method (both fifth and sixth rows) in comparison to the baseline model. However, the average difference between baseline and proposed segmentation performance is more evident for the NuInsSeg dataset (4.9\%, 5.4\%, and 5.9\% for Dice, AJI, and PQ score, respectively). The proposed method with TTSN also outperforms the other applied normalization techniques (3.4\% for Dice, 3.1\% for AJI, and 2.5\% for PQ score, respectively for the offline normalization method, and 6.8\% for Dice, 6.4\% for AIJ and 5.9\% for PQ score, respectively for the extended offline normalization strategy, and 1.7\% for Dice, 2.8\% for AIJ and 2.7\% for PQ score, respectively for the offline normalization with atlas image approach). For qualitative comparison, we show sample segmentation results from the baseline model and the proposed method in Fig.~\ref{fig:compare_res}.

\begin{table}
	\centering
	\caption{Segmentation results for the NuInsSeg test data (average and standard deviation over three runs). The last two rows represent the results from the proposed approach. Non-det.: Non-deterministic; TTSN: Test Time Stain Normalization; AJI: Aggregate Jaccard Index; PQ: Panoptic Quality.}\label{nuinsseg_res}
	\begin{tabular}{l|c|c|c}
		\hline
		Method &  Dice (\%) & \ AJI (\%) & PQ (\%)   \\
		\hline
		No normalization (baseline)                   & 66.6 $\pm$ 1.0 & 42.1 $\pm$ 0.4 & 35.0 $\pm$ 0.3 \\
		Offline normalization                         & 68.1 $\pm$ 1.2 & 44.4 $\pm$ 1.0 & 38.4 $\pm$ 0.9\\
		Extended offline normalization                & 64.7 $\pm$ 0.8 & 41.1 $\pm$ 0.1 & 35.0 $\pm$ 1.6 \\
		Offline normalization with atlas              & 69.8 $\pm$ 0.8  & 44.7 $\pm$ 0.6 & 38.2 $\pm$ 0.5\\\hline
		Non-det. stain normalization                  & 69.8 $\pm$ 0.9 & 45.7 $\pm$ 0.7 & 39.5 $\pm$ 0.7  \\
		Non-det. stain normalization + TTSN           & 71.5 $\pm$ 0.4 & 47.5 $\pm$ 0.4 & 40.9 $\pm$ 0.6 \\
		
		\hline
	\end{tabular}
\end{table}

\begin{table}[]
	\centering
	\caption{Comparison between incorporating 7 or 14 images for deterministic stain normalization in the inference phase. TTSN: Test Time Stain Normalization; AJI: Aggregate Jaccard Index; PQ: Panoptic Quality.}\label{images14}
	\begin{tabular}{l|cc|cc|cc}
		\hline
		\multirow{2}{*}{dataset} & \multicolumn{2}{c|}{Dice (\%)}                & \multicolumn{2}{c|}{AJI (\%)}                  & \multicolumn{2}{c}{PQ (\%)}                    \\ \cline{2-7} 
		& \multicolumn{1}{c|}{TTSN (7)} & TTSN(14) & \multicolumn{1}{c|}{TTSN (7)} & TTSN (14) & \multicolumn{1}{c|}{TTSN (7)} & TTSN (14) \\ \hline
		MoNuseg      & \multicolumn{1}{c|}{83.40$\pm$0.07} &  83.42$\pm$0.04   & \multicolumn{1}{c|}{66.91$\pm$0.21}   &   66.93$\pm$0.20   & \multicolumn{1}{c|}{64.89$\pm$0.32}   & 64.87$\pm$0.31    \\
		TNBC         & \multicolumn{1}{c|}{81.36$\pm$0.19}  &  81.40$\pm$0.17   & \multicolumn{1}{c|}{64.78$\pm$0.24}   &   64.83$\pm$0.25    & \multicolumn{1}{c|}{62.27$\pm$0.41}   & 62.27$\pm$0.38     \\
		CryoNuSeg    & \multicolumn{1}{c|}{79.04$\pm$0.18}  &  79.04$\pm$0.18   & \multicolumn{1}{c|}{52.01$\pm$0.10}   &   52.04$\pm$0.13    & \multicolumn{1}{c|}{49.17$\pm$0.24}   & 49.16$\pm$0.24      \\
		CPM-15       & \multicolumn{1}{c|}{75.74$\pm$0.62}  &  75.74$\pm$0.62   & \multicolumn{1}{c|}{54.65$\pm$0.57}   &   54.64$\pm$0.56    & \multicolumn{1}{c|}{50.85$\pm$0.93}   & 50.82$\pm$0.96      \\
		CMP-17       & \multicolumn{1}{c|}{82.28$\pm$0.39}  &  82.27$\pm$0.39   & \multicolumn{1}{c|}{64.35$\pm$0.64}   &   64.35$\pm$0.63    & \multicolumn{1}{c|}{59.67$\pm$0.79}   & 59.70$\pm$0.79      \\
		ConSep       & \multicolumn{1}{c|}{71.51$\pm$0.08}  &  71.52$\pm$0.09   & \multicolumn{1}{c|}{42.68$\pm$0.26}   &   42.69$\pm$0.25    & \multicolumn{1}{c|}{41.85$\pm$0.14}   & 41.87$\pm$0.13       \\
		NuInsSeg     & \multicolumn{1}{c|}{71.50$\pm$0.38}  &  71.53$\pm$0.36   & \multicolumn{1}{c|}{47.47$\pm$0.39}   &   47.49$\pm$0.37    & \multicolumn{1}{c|}{40.85$\pm$0.65}   & 40.81$\pm$0.67       \\ \hline
	\end{tabular}
\end{table}

The derived standard deviations of the results indicate that the obtained outcomes from the proposed model with TTSN are highly robust, with an average standard deviation of 0.28\%, 0.34\%, and 0.47\% for Dice, AJI, and PQ scores, respectively. Similarly, the model remains robust even without TTSN, with an average standard deviation of 0.41\%, 0.48\%, and 0.6\% for Dice, AJI, and PQ scores, respectively.

There are some limitations in this study that can be addressed in future works. First, while we propose a generalizable framework for the nuclei instance segmentation task, we only used one of the state-of-the-art DL-based models (DDU-Net) in our study. Although DDU-Net has shown excellent nuclei instance segmentation performance~\cite{9745980, 10.3389/fmed.2022.978146, 9446924}, DDU-Net can be replaced by other state-of-the-art segmentation models in future research. However, we would like to emphasize that most other state-of-the-art DL-based models for nuclei instance segmentation (such as triple U-Net, HoverNet, or attention augmented distance regression model~\cite{ZHAO2020101786, graham2019hover, DOGAR2023104199}) have similar encoder-decoder-based architectures to the utilized DDU-Net model. Second, in this study, we focused on the nuclei instance segmentation task, but the proposed framework can be evaluated for nuclei detection or nuclei instance segmentation and classification or nuclei detection tasks in the future as well. Finally, using stain normalization, as shown in former studies~\cite{jahanifar2023domain, li2023laplacian}, introduces computational overhead, especially in the inference phase. While we observe improved nuclei instance performance in all test datasets, the gains in some datasets were not as notable as in other datasets. Thus, the application of the proposed TTSN method to increase the generalization at the expense of extra computation overhead should be considered, especially when limited computational resources are available.

\section{Conclusion}
While many ML-based and DL-based approaches have been proposed for nuclei instance segmentation in histological images, their performance usually degrades when tested on unseen new images. We proposed a framework for generalized nuclei instance segmentation with non-deterministic train time and deterministic test time stain normalization. Applied on seven independent test datasets, the results showed the superior performance of the proposed method compared to the baseline segmentation model. Therefore, the proposed approach can be considered a generalized framework for nuclei instance segmentation.

\subsubsection{Acknowledgements} We thank NVIDIA corporation for the generous GPU donation. This project was partially supported by the Austrian Research Promotion Agency (FFG), No. 872636.

\subsubsection{Competing interests}
The authors declare no competing interests.

\bibliographystyle{splncs04}
\bibliography{gen_seg.bib}

\begin{thebibliography}{10}
\providecommand{\url}[1]{\texttt{#1}}
\providecommand{\urlprefix}{URL }
\providecommand{\doi}[1]{https://doi.org/#1}

\bibitem{anwar2018medical}
Anwar, S.M., Majid, M., Qayyum, A., Awais, M., Alnowami, M., Khan, M.K.:
  Medical image analysis using convolutional neural networks: A review. Journal
  of Medical Systems  \textbf{42}(11), ~226 (2018).
  \doi{10.1007/s10916-018-1088-1},
  \url{https://doi.org/10.1007/s10916-018-1088-1}

\bibitem{aresta2019bach}
Aresta, G., Araújo, T., Kwok, S., Chennamsetty, S.S., Safwan, M., Alex, V.,
  Marami, B., Prastawa, M., Chan, M., Donovan, M., Fernandez, G., Zeineh, J.,
  Kohl, M., Walz, C., Ludwig, F., Braunewell, S., Baust, M., Vu, Q.D., To,
  M.N.N., Kim, E., Kwak, J.T., Galal, S., Sanchez-Freire, V., Brancati, N.,
  Frucci, M., Riccio, D., Wang, Y., Sun, L., Ma, K., Fang, J., Kone, I.,
  Boulmane, L., Campilho, A., Eloy, C., Polónia, A., Aguiar, P.: {BACH}: Grand
  challenge on breast cancer histology images. Medical Image Analysis
  \textbf{56},  122--139 (2019).
  \doi{https://doi.org/10.1016/j.media.2019.05.010}

\bibitem{pmlr-v156-bancher21a}
Bancher, B., Mahbod, A., Ellinger, I., Ecker, R., Dorffner, G.: Improving {Mask
  R-CNN} for nuclei instance segmentation in hematoxylin \& eosin-stained
  histological images. In: MICCAI Workshop on Computational Pathology.
  vol.~156, pp. 20--35 (2021)

\bibitem{Basu2023}
Basu, A., Senapati, P., Deb, M., Rai, R., Dhal, K.G.: A survey on recent trends
  in deep learning for nucleus segmentation from histopathology images.
  Evolving Systems  (2023). \doi{10.1007/s12530-023-09491-3}

\bibitem{9622821}
Brady, K., Khorrami, P., Gjesteby, L., Brattain, L.: Instance segmentation of
  neuronal nuclei leveraging domain adaptation. In: IEEE High Performance
  Extreme Computing Conference. pp.~1--5 (2021).
  \doi{10.1109/HPEC49654.2021.9622821}

\bibitem{chen2017dcan}
Chen, H., Qi, X., Yu, L., Dou, Q., Qin, J., Heng, P.A.: {DCAN}: Deep
  contour-aware networks for object instance segmentation from histology
  images. Medical Image Analysis  \textbf{36},  135--146 (2017).
  \doi{https://doi.org/10.1016/j.media.2016.11.004}

\bibitem{DOGAR2023104199}
Dogar, G.M., Shahzad, M., Fraz, M.M.: Attention augmented distance regression
  and classification network for nuclei instance segmentation and type
  classification in histology images. Biomedical Signal Processing and Control
  \textbf{79},  104199 (2023). \doi{https://doi.org/10.1016/j.bspc.2022.104199}

\bibitem{fischer2020nuclear}
Fischer, E.G.: Nuclear morphology and the biology of cancer cells. Acta
  cytologica  \textbf{64}(6),  511--519 (2020). \doi{10.1159/000508780}

\bibitem{9745980}
Foucart, A., Debeir, O., Decaestecker, C.: Comments on “{MoNuSAC2020}: A
  multi-organ nuclei segmentation and classification challenge”. IEEE
  Transactions on Medical Imaging  \textbf{41}(4),  997--999 (2022).
  \doi{10.1109/TMI.2022.3156023}

\bibitem{GEORGE2020103954}
George, K., Faziludeen, S., Sankaran, P., {Joseph K}, P.: Breast cancer
  detection from biopsy images using nucleus guided transfer learning and
  belief based fusion. Computers in Biology and Medicine  \textbf{124},  103954
  (2020). \doi{https://doi.org/10.1016/j.compbiomed.2020.103954}

\bibitem{graham2019hover}
Graham, S., Vu, Q.D., Raza, S.E.A., Azam, A., Tsang, Y.W., Kwak, J.T., Rajpoot,
  N.: {Hover-Net}: Simultaneous segmentation and classification of nuclei in
  multi-tissue histology images. Medical Image Analysis  \textbf{58},  101563
  (2019). \doi{10.1016/j.media.2019.101563}

\bibitem{Hameed2022}
Hameed, Z., Garcia-Zapirain, B., Aguirre, J.J., Isaza-Ruget, M.A.: Multiclass
  classification of breast cancer histopathology images using multilevel
  features of deep convolutional neural network. Scientific Reports
  \textbf{12}(1),  15600 (2022). \doi{10.1038/s41598-022-19278-2}

\bibitem{haq2023nusegda}
Haq, M.M., Ma, H., Huang, J.: {NuSegDA}: Domain adaptation for nuclei
  segmentation. Frontiers in Big Data  \textbf{6},  1108659 (2023).
  \doi{https://doi.org/10.3389/fdata.2023.1108659}

\bibitem{HOLLANDI2022295}
Hollandi, R., Moshkov, N., Paavolainen, L., Tasnadi, E., Piccinini, F.,
  Horvath, P.: Nucleus segmentation: towards automated solutions. Trends in
  Cell Biology  \textbf{32}(4),  295--310 (2022).
  \doi{10.1016/j.tcb.2021.12.004}

\bibitem{irshad2014methods}
Irshad, H., Veillard, A., Roux, L., Racoceanu, D.: Methods for nuclei
  detection, segmentation, and classification in digital histopathology: A
  review—current status and future potential. IEEE Reviews in Biomedical
  Engineering  \textbf{7},  97--114 (2014). \doi{10.1109/RBME.2013.2295804}

\bibitem{jahanifar2023domain}
Jahanifar, M., Raza, M., Xu, K., Vuong, T., Jewsbury, R., Shephard, A.,
  Zamanitajeddin, N., Kwak, J.T., Raza, S.E.A., Minhas, F., et~al.: Domain
  generalization in computational pathology: Survey and guidelines. arXiv
  preprint arXiv:2310.19656  (2023)

\bibitem{johnson2018adapting}
Johnson, J.W.: Adapting {Mask-RCNN} for automatic nucleus segmentation. arXiv
  preprint arXiv:1805.00500  (2018)

\bibitem{Jung2019}
Jung, H., Lodhi, B., Kang, J.: An automatic nuclei segmentation method based on
  deep convolutional neural networks for histopathology images. BMC Biomedical
  Engineering  \textbf{1}(1), ~24 (2019). \doi{10.1186/s42490-019-0026-8}

\bibitem{khan2014nonlinear}
Khan, A.M., Rajpoot, N., Treanor, D., Magee, D.: A nonlinear mapping approach
  to stain normalization in digital histopathology images using image-specific
  color deconvolution. IEEE Transactions on Biomedical Engineering
  \textbf{61}(6),  1729--1738 (2014). \doi{10.1109/TBME.2014.2303294}

\bibitem{Kingma2014}
Kingma, D.P., Ba, J.: {Adam}: {A} method for stochastic optimization. In:
  International Conference on Learning Representations (2015)

\bibitem{Kirillov_2019_CVPR}
Kirillov, A., He, K., Girshick, R., Rother, C., Dollar, P.: Panoptic
  segmentation. In: Conference on Computer Vision and Pattern Recognition. pp.
  9404--9413 (2019)

\bibitem{Kumar2017}
Kumar, N., Verma, R., Sharma, S., Bhargava, S., Vahadane, A., Sethi, A.: A
  dataset and a technique for generalized nuclear segmentation for
  computational pathology. IEEE Transactions on Medical Imaging
  \textbf{36}(7),  1550--1560 (2017). \doi{10.1109/TMI.2017.2677499}

\bibitem{monuseg}
Kumar, N., Verma, R., Anand, D., Zhou, Y., Onder, O.F., Tsougenis, E., Chen,
  H., Heng, P.A., Li, J., Hu, Z., Wang, Y., Koohbanani, N.A., Jahanifar, M.,
  Tajeddin, N.Z., Gooya, A., Rajpoot, N., Ren, X., Zhou, S., Wang, Q., Shen,
  D., Yang, C.K., Weng, C.H., Yu, W.H., Yeh, C.Y., Yang, S., Xu, S., Yeung,
  P.H., Sun, P., Mahbod, A., Schaefer, G., Ellinger, I., Ecker, R., Smedby, O.,
  Wang, C., Chidester, B., Ton, T.V., Tran, M.T., Ma, J., Do, M.N., Graham, S.,
  Vu, Q.D., Kwak, J.T., Gunda, A., Chunduri, R., Hu, C., Zhou, X., Lotfi, D.,
  Safdari, R., Kascenas, A., O’Neil, A., Eschweiler, D., Stegmaier, J., Cui,
  Y., Yin, B., Chen, K., Tian, X., Gruening, P., Barth, E., Arbel, E., Remer,
  I., Ben-Dor, A., Sirazitdinova, E., Kohl, M., Braunewell, S., Li, Y., Xie,
  X., Shen, L., Ma, J., Baksi, K.D., Khan, M.A., Choo, J., Colomer, A.,
  Naranjo, V., Pei, L., Iftekharuddin, K.M., Roy, K., Bhattacharjee, D.,
  Pedraza, A., Bueno, M.G., Devanathan, S., Radhakrishnan, S., Koduganty, P.,
  Wu, Z., Cai, G., Liu, X., Wang, Y., Sethi, A.: A multi-organ nucleus
  segmentation challenge. IEEE Transactions on Medical Imaging  \textbf{39}(5),
   1380--1391 (2020). \doi{10.1109/TMI.2019.2947628}

\bibitem{10.1007/978-3-031-16449-1_68}
Li, C., Liu, D., Li, H., Zhang, Z., Lu, G., Chang, X., Cai, W.: Domain adaptive
  nuclei instance segmentation and classification via category-aware feature
  alignment and pseudo-labelling. In: Wang, L., Dou, Q., Fletcher, P.T.,
  Speidel, S., Li, S. (eds.) Medical Image Computing and Computer Assisted
  Intervention -- MICCAI 2022. pp. 715--724. Springer Nature Switzerland, Cham
  (2022). \doi{https://doi.org/10.1007/978-3-031-16449-1_68}

\bibitem{li2023laplacian}
Li, F., Hu, Z., Chen, W., Kak, A.: A laplacian pyramid based generative h\&e
  stain augmentation network. IEEE Transactions on Medical Imaging pp.~1--1
  (2023). \doi{10.1109/TMI.2023.3317239}

\bibitem{9156759}
Liu, D., Zhang, D., Song, Y., Zhang, F., ODonnell, L., Huang, H., Chen, M.,
  Cai, W.: Unsupervised instance segmentation in microscopy images via panoptic
  domain adaptation and task re-weighting. In: IEEE/CVF Conference on Computer
  Vision and Pattern Recognition. pp. 4242--4251 (2020).
  \doi{10.1109/CVPR42600.2020.00430}

\bibitem{macenko2009method}
Macenko, M., Niethammer, M., Marron, J.S., Borland, D., Woosley, J.T., Guan,
  X., Schmitt, C., Thomas, N.E.: A method for normalizing histology slides for
  quantitative analysis. In: International Symposium on Biomedical Imaging. pp.
  1107--1110 (2009). \doi{10.1109/ISBI.2009.5193250}

\bibitem{mahbod2022deep}
Mahbod, A., Entezari, R., Ellinger, I., Saukh, O.: Deep neural network pruning
  for nuclei instance segmentation in hematoxylin and eosin-stained
  histological images. In: Applications of Medical Artificial Intelligence. pp.
  108--117. Springer Nature Switzerland (2022).
  \doi{10.1007/978-3-031-17721-7_12}

\bibitem{mahbod2023nuinsseg}
Mahbod, A., Polak, C., Feldmann, K., Khan, R., Gelles, K., Dorffner, G.,
  Woitek, R., Hatamikia, S., Ellinger, I.: Nuinsseg: A fully annotated dataset
  for nuclei instance segmentation in h\&e-stained histological images. arXiv
  preprint arXiv:2308.01760  (2023)

\bibitem{mahbod2021cryonuseg_org}
Mahbod, A., Schaefer, G., Bancher, B., Löw, C., Dorffner, G., Ecker, R.,
  Ellinger, I.: {CryoNuSeg: A dataset for nuclei instance segmentation of
  cryosectioned H\&E-stained histological images}. Computers in Biology and
  Medicine  \textbf{132},  104349 (2021).
  \doi{10.1016/j.compbiomed.2021.104349}

\bibitem{10.3389/fmed.2022.978146}
Mahbod, A., Schaefer, G., Dorffner, G., Hatamikia, S., Ecker, R., Ellinger, I.:
  A dual decoder u-net-based model for nuclei instance segmentation in
  hematoxylin and eosin-stained histological images. Frontiers in Medicine
  \textbf{9} (2022). \doi{10.3389/fmed.2022.978146}

\bibitem{mahbod2021pollen}
Mahbod, A., Schaefer, G., Ecker, R., Ellinger, I.: Pollen grain microscopic
  image classification using an ensemble of fine-tuned deep convolutional
  neural networks. In: International Conference on Pattern Recognition. pp.
  344--356. Springer (2021). \doi{10.1007/978-3-030-68763-2_26}

\bibitem{10.1007/978-3-030-23937-4_9}
Mahbod, A., Schaefer, G., Ellinger, I., Ecker, R., Smedby, {\"O}., Wang, C.: A
  two-stage {U-Net} algorithm for segmentation of nuclei in {H{\&}E}-stained
  tissues. In: European Congress on Digital Pathology. pp. 75--82 (2019).
  \doi{10.1007/978-3-030-23937-4_9}

\bibitem{https://doi.org/10.1111/tbj.13728}
Masood, S.: The changing role of pathologists from morphologists to molecular
  pathologists in the era of precision medicine. The Breast Journal
  \textbf{26}(1),  27--34 (2020). \doi{https://doi.org/10.1111/tbj.13728}

\bibitem{meijering2012cell}
Meijering, E.: Cell segmentation: 50 years down the road [life sciences]. IEEE
  Signal Processing Magazine  \textbf{29}(5),  140--145 (2012).
  \doi{10.1109/MSP.2012.2204190}

\bibitem{diagnostics13142379}
Moscalu, M., Moscalu, R., Dascălu, C.G., Țarcă, V., Cojocaru, E., Costin,
  I.M., Țarcă, E., Șerban, I.L.: Histopathological images analysis and
  predictive modeling implemented in digital pathology\&mdash;current affairs
  and perspectives. Diagnostics  \textbf{13}(14) (2023).
  \doi{10.3390/diagnostics13142379}

\bibitem{naylor2018segmentation}
{Naylor}, P., {La\'e}, M., {Reyal}, F., {Walter}, T.: Segmentation of nuclei in
  histopathology images by deep regression of the distance map. IEEE
  Transactions on Medical Imaging  \textbf{38}(2),  448--459 (Feb 2019).
  \doi{10.1109/TMI.2018.2865709}

\bibitem{10.3389/fbioe.2019.00300}
Pontalba, J.T., Gwynne-Timothy, T., David, E., Jakate, K., Androutsos, D.,
  Khademi, A.: Assessing the impact of color normalization in convolutional
  neural network-based nuclei segmentation frameworks. Frontiers in
  Bioengineering and Biotechnology  \textbf{7} (2019).
  \doi{10.3389/fbioe.2019.00300}

\bibitem{reinhard2001color}
Reinhard, E., Adhikhmin, M., Gooch, B., Shirley, P.: Color transfer between
  images. IEEE Computer Graphics and Applications  \textbf{21}(5),  34--41
  (2001). \doi{10.1109/38.946629}

\bibitem{Ronneberger2015}
Ronneberger, O., Fischer, P., Brox, T.: {U-Net}: Convolutional networks for
  biomedical image segmentation. In: International Conference on Medical Image
  Computing and Computer-Assisted Intervention. pp. 234--241 (2015).
  \doi{10.1007/978-3-319-24574-4_28}

\bibitem{SALVI2021104129}
Salvi, M., Acharya, U.R., Molinari, F., Meiburger, K.M.: The impact of pre- and
  post-image processing techniques on deep learning frameworks: A comprehensive
  review for digital pathology image analysis. Computers in Biology and
  Medicine  \textbf{128},  104129 (2021).
  \doi{https://doi.org/10.1016/j.compbiomed.2020.104129}

\bibitem{SHAMSHAD2023102802}
Shamshad, F., Khan, S., Zamir, S.W., Khan, M.H., Hayat, M., Khan, F.S., Fu, H.:
  Transformers in medical imaging: A survey. Medical Image Analysis
  \textbf{88},  102802 (2023).
  \doi{https://doi.org/10.1016/j.media.2023.102802}

\bibitem{skinner2017nuclear}
Skinner, B.M., Johnson, E.E.: Nuclear morphologies: their diversity and
  functional relevance. Chromosoma  \textbf{126}(2),  195--212 (2017).
  \doi{10.1007/s00412-016-0614-5}

\bibitem{tellez2019quantifying}
Tellez, D., Litjens, G., Bándi, P., Bulten, W., Bokhorst, J.M., Ciompi, F.,
  {van der Laak}, J.: Quantifying the effects of data augmentation and stain
  color normalization in convolutional neural networks for computational
  pathology. Medical Image Analysis  \textbf{58},  101544 (2019).
  \doi{https://doi.org/10.1016/j.media.2019.101544}

\bibitem{7164042}
Vahadane, A., Peng, T., Albarqouni, S., Baust, M., Steiger, K., Schlitter,
  A.M., Sethi, A., Esposito, I., Navab, N.: Structure-preserved color
  normalization for histological images. In: 12th International Symposium on
  Biomedical Imaging. pp. 1012--1015 (2015). \doi{10.1109/ISBI.2015.7164042}

\bibitem{VASILJEVIC2023110780}
Vasiljević, J., Feuerhake, F., Wemmert, C., Lampert, T.: {HistoStarGAN:} a
  unified approach to stain normalisation, stain transfer and stain invariant
  segmentation in renal histopathology. Knowledge-Based Systems  \textbf{277},
  110780 (2023). \doi{https://doi.org/10.1016/j.knosys.2023.110780}

\bibitem{VASILJEVIC2022102420}
Vasiljević, J., Nisar, Z., Feuerhake, F., Wemmert, C., Lampert, T.: {CycleGAN}
  for virtual stain transfer: Is seeing really believing? Artificial
  Intelligence in Medicine  \textbf{133},  102420 (2022).
  \doi{https://doi.org/10.1016/j.artmed.2022.102420}

\bibitem{9446924}
Verma, R., Kumar, N., Patil, A., Kurian, N.C., Rane, S., Graham, S., Vu, Q.D.,
  Zwager, M., Raza, S.E.A., Rajpoot, N., Wu, X., Chen, H., Huang, Y., Wang, L.,
  Jung, H., Brown, G.T., Liu, Y., Liu, S., Jahromi, S.A.F., Khani, A.A.,
  Montahaei, E., Baghshah, M.S., Behroozi, H., Semkin, P., Rassadin, A.,
  Dutande, P., Lodaya, R., Baid, U., Baheti, B., Talbar, S., Mahbod, A., Ecker,
  R., Ellinger, I., Luo, Z., Dong, B., Xu, Z., Yao, Y., Lv, S., Feng, M., Xu,
  K., Zunair, H., Hamza, A.B., Smiley, S., Yin, T.K., Fang, Q.R., Srivastava,
  S., Mahapatra, D., Trnavska, L., Zhang, H., Narayanan, P.L., Law, J., Yuan,
  Y., Tejomay, A., Mitkari, A., Koka, D., Ramachandra, V., Kini, L., Sethi, A.:
  {MoNuSAC2020}: A multi-organ nuclei segmentation and classification
  challenge. IEEE Transactions on Medical Imaging pp.~1--1 (2021).
  \doi{10.1109/TMI.2021.3085712}

\bibitem{9745890}
Verma, R., Kumar, N., Patil, A., Kurian, N.C., Rane, S., Sethi, A.: Author’s
  reply to “{MoNuSAC2020}: A multi-organ nuclei segmentation and
  classification challenge”. IEEE Transactions on Medical Imaging
  \textbf{41}(4),  1000--1003 (2022). \doi{10.1109/TMI.2022.3157048}

\bibitem{Voon2023}
Voon, W., Hum, Y.C., Tee, Y.K., Yap, W.S., Nisar, H., Mokayed, H., Gupta, N.,
  Lai, K.W.: Evaluating the effectiveness of stain normalization techniques in
  automated grading of invasive ductal carcinoma histopathological images.
  Scientific Reports  \textbf{13}(1),  20518 (2023).
  \doi{10.1038/s41598-023-46619-6}

\bibitem{vu2019methods}
Vu, Q.D., Graham, S., Kurc, T., To, M.N.N., Shaban, M., Qaiser, T., Koohbanani,
  N.A., Khurram, S.A., Kalpathy-Cramer, J., Zhao, T., et~al.: Methods for
  segmentation and classification of digital microscopy tissue images.
  Frontiers in Bioengineering and Biotechnology  \textbf{7}, ~53 (2019).
  \doi{10.3389/fbioe.2019.00053}

\bibitem{10.1007/978-3-031-21014-3_26}
Vu, Q.D., Jewsbury, R., Graham, S., Jahanifar, M., Raza, S.E.A., Minhas, F.,
  Bhalerao, A., Rajpoot, N.: Nuclear segmentation and classification: On color
  and compression generalization. In: Lian, C., Cao, X., Rekik, I., Xu, X.,
  Cui, Z. (eds.) Machine Learning in Medical Imaging. pp. 249--258. Springer
  Nature Switzerland, Cham (2022)

\bibitem{8759574}
{Vuola}, A.O., {Akram}, S.U., {Kannala}, J.: {Mask-RCNN} and {U-Net} ensembled
  for nuclei segmentation. In: International Symposium on Biomedical Imaging.
  pp. 208--212 (2019). \doi{10.1109/ISBI.2019.8759574}

\bibitem{wang2022fuseg}
Wang, C., Mahbod, A., Ellinger, I., Galdran, A., Gopalakrishnan, S., Niezgoda,
  J., Yu, Z.: {FUSeg}: The foot ulcer segmentation challenge. arXiv preprint
  arXiv:2201.00414  (2022)

\bibitem{WANG2023107768}
Wang, S., Rong, R., Gu, Z., Fujimoto, J., Zhan, X., Xie, Y., Xiao, G.:
  Unsupervised domain adaptation for nuclei segmentation: Adapting from
  hematoxylin \& eosin stained slides to immunohistochemistry stained slides
  using a curriculum approach. Computer Methods and Programs in Biomedicine
  \textbf{241},  107768 (2023).
  \doi{https://doi.org/10.1016/j.cmpb.2023.107768}

\bibitem{7373572}
Xing, F., Yang, L.: Robust nucleus/cell detection and segmentation in digital
  pathology and microscopy images: A comprehensive review. IEEE Reviews in
  Biomedical Engineering  \textbf{9},  234--263 (2016).
  \doi{10.1109/RBME.2016.2515127}

\bibitem{10190115}
Xu, H., Xu, Q., Cong, F., Kang, J., Han, C., Liu, Z., Madabhushi, A., Lu, C.:
  Vision transformers for computational histopathology. IEEE Reviews in
  Biomedical Engineering pp. 1--17 (2023). \doi{10.1109/RBME.2023.3297604}

\bibitem{ZHAO2020101786}
Zhao, B., Chen, X., Li, Z., Yu, Z., Yao, S., Yan, L., Wang, Y., Liu, Z., Liang,
  C., Han, C.: Triple {U-net}: Hematoxylin-aware nuclei segmentation with
  progressive dense feature aggregation. Medical Image Analysis  \textbf{65},
  101786 (2020). \doi{10.1016/j.media.2020.101786}

\bibitem{8237506}
Zhu, J.Y., Park, T., Isola, P., Efros, A.A.: Unpaired image-to-image
  translation using cycle-consistent adversarial networks. In: IEEE
  International Conference on Computer Vision. pp. 2242--2251 (2017).
  \doi{10.1109/ICCV.2017.244}

\end{thebibliography}

\end{document}